\documentclass[twocolumn,secnumarabic,amssymb, nobibnotes, aps, prl, superscriptaddress, nobalancelastpage,longbibliography,floatfix]{revtex4-1}

 \usepackage{graphicx}
\usepackage{bm}
\usepackage{amsmath} \usepackage{braket} \usepackage{epsfig}
\usepackage{tensor}
\usepackage{multirow}
\usepackage[T1]{fontenc}
\usepackage[version=4]{mhchem}
\usepackage{titlesec}
\usepackage{ifthen}
\usepackage{sidecap}
\usepackage{listings}
\usepackage[para,online,flushleft]{threeparttablex}
\usepackage{algorithm}
\usepackage{algpseudocode}
\usepackage{xr-hyper}
\usepackage{hyperref}
\hypersetup{breaklinks=true,colorlinks=true,linkcolor=blue,citecolor=blue,filecolor=magenta,urlcolor=cyan}

\usepackage[all]{hypcap}

\renewcommand{\vec}[1]{\mbox{\boldmath $#1$}}
\newcommand{\PRLsep}{\noindent\makebox[\linewidth]{\resizebox{0.3333\linewidth}{1pt}{$\bullet$}}\bigskip}

\usepackage[normalem]{ulem}

\usepackage{xcolor}
\definecolor{pastelgray}{rgb}{0.81, 0.81, 0.77}
\definecolor{beaublue}{rgb}{0.9, 0.9, 0.93}

\makeatletter
\def\@bibdataout@aps{%
\immediate\write\@bibdataout{%
@CONTROL{%
apsrev41Control%
\longbibliography@sw{%
    ,author="08",editor="1",pages="1",title="0",year="1"%
    }{%
    ,author="08",editor="1",pages="1",title="",year="1"%
    }%
  }%
}%
\if@filesw \immediate \write \@auxout {\string \citation {apsrev41Control}}\fi
}
\makeatother

\begin{document}

\title{Nudged elastic band approach to nuclear fission pathways}

\author{Eric Flynn}
\affiliation{Department of Physics and Astronomy and FRIB/NSCL Laboratory, Michigan State University, East Lansing, Michigan 48824, USA}

\author{Daniel Lay}
\affiliation{Department of Physics and Astronomy and FRIB/NSCL Laboratory, Michigan State University, East Lansing, Michigan 48824, USA}

\author{Sylvester Agbemava}
\affiliation{FRIB/NSCL Laboratory, Michigan State University, East Lansing, Michigan 48824, USA}

\author{Pablo Giuliani}
\affiliation{FRIB/NSCL Laboratory, Michigan State University, East Lansing, Michigan 48824, USA}
\affiliation{Department of Statistics and Probability, Michigan State University, East Lansing, Michigan 48824, USA}

\author{Kyle Godbey}
\affiliation{FRIB/NSCL Laboratory, Michigan State University, East Lansing, Michigan 48824, USA}

\author{Witold Nazarewicz}
\affiliation{Department of Physics and Astronomy and FRIB/NSCL Laboratory, Michigan State University, East Lansing, Michigan 48824, USA}

\author{Jhilam Sadhukhan}
\affiliation{Physics Group, Variable Energy Cyclotron Centre, Kolkata 700064, India }
\affiliation{Homi Bhabha National Institute, Mumbai 400094, India }

\begin{abstract}
\edef\oldrightskip{\the\rightskip}
\begin{description}
\rightskip\oldrightskip\relax
\setlength{\parskip}{0pt}
\item[Background]
The nuclear fission process is a dramatic example of the  large-amplitude collective motion in which the  nucleus undergoes a series of shape changes  before splitting into distinct fragments. This motion can be represented by a pathway in the many-dimensional  space of collective coordinates. The collective action along the fission pathway determines the spontaneous fission half-lives as well as  mass and charge distributions of fission fragments.

\item[Purpose]
We study the performance and precision of various methods to determine the minimum-action and minimum-energy  fission trajectories in the collective space.

\item[Methods]
We apply the  nudged elastic band  method (NEB), grid-based methods, and Euler-Lagrange approach to the collective action minimization  in two- and three-dimensional collective spaces. 

\item[Results]
The performance of various  approaches to the fission pathway problem is assessed  by studying the collective motion along both analytic energy surfaces and realistic potential energy surfaces obtained with the  Skyrme-Hartree-Fock-Bogoliubov theory. The uniqueness and stability of the solutions is studied. The NEB  method is capable of efficient determination of the exit points on the outer turning surface that  characterize the most probable fission pathway and constitute the key input for fission studies.
This method can also be used to accurately compute  the critical points (i.e., local minima and saddle points) on the potential energy surface of the fissioning nucleus that determine the static fission path. The dynamic programming method also performs quite well and it can be used in many-dimensional cases to provide initial conditions for the NEB calculations.

\item[Conclusions]

The  NEB method is the tool of choice for finding the least-action and minimum-energy fission trajectories. It will be particularly  useful in large-scale fission calculation 
of superheavy nuclei and neutron-rich fissioning nuclei contributing to the astrophysical r-process recycling.

\end{description}
\end{abstract}

\date{\today}

\maketitle

\section{Introduction}
\label{sec:introduction}
Fission is a fundamental nuclear decay that is important in many areas of science, ranging from structure and stability of heavy and superheavy nuclei \cite{krappe2012,Schmidt2018,GiulianiRMP} to studies devoted to physics beyond the standard model of particle physics~\cite{vogel2015} and the synthesis of heavy elements~\cite{horowitz2018,vassh2019,Giuliani2020}. 

Theoretically, the nuclear fission process is an example of the nuclear large-amplitude collective motion originating from the single-particle motion of individual nucleons. Due to the complexity of this process, our understanding of nuclear fission is still incomplete.
For the  state of affairs in this field, we refer to the recent review \cite{schunck2016,Bender2020}.

When it comes to realistic predictions, the self-consistent nuclear energy density functional (EDF) method \cite{BenderRMP,SchunckBook} has proven to be very
successful in terms of quantitative reproduction of fission lifetimes and fragment yields.
Unfortunately, realistic self-consistent fission calculations in a multidimensional
collective space, based on the microscopic
input, are computationally expensive when it comes to large-scale theoretical
fission surveys.  Given the
computational cost of microscopic methods and the large number of fissioning
nuclei that are, e.g.,  expected to contribute to the astrophysical r-process nucleosynthesis, 
calculations have mostly relied on simple parametrizations or highly
phenomenological models. The new perspective is offered by
 state-of-the-art theoretical frameworks and  modern computational techniques that promise  to speed up the  calculations to be able to carry out quantified global fission surveys for multiple inputs \cite{Bender2020}.
 
This study is concerned with finding the optimal pathway during the tunneling motion phase of spontaneous fission (SF). Such a trajectory, dubbed the least-action path (LAP), is obtained by minimizing the collective action in a many-dimensional collective space \cite{Kapur1937,Brack1972}. A number of techniques have been proposed to deal with this challenging task. In the early application
\cite{Ledergerber1973}, the trial pathways were assumed in a parametrized  form and the LAP was obtained by  minimizing the penetration integral with respect to the variational parameters.
Grid-based techniques such as the dynamic-programming~\cite{Baran1981} and Ritz~\cite{Baran1978} methods have been used in numerous EDF calculations of LAPs
\cite{Sadhukhan2013,Sadhukhan2014,Sadhukhan2016,Sadhukhan2017,Zhao2015,Zhao2016,Mercier2021}. In Refs.~\cite{Schmid1986,Eckern1992,Kindo1989,Iwamoto1992,Iwamoto1994,Scamps2015} LAPs were obtained by
solving the eikonal
equation by the method of characteristics. Effectively, this method  can be related to a quantum mechanical propagation in imaginary time that amounts to  solving the
classical equations of motion in an inverted potential. Within this approach,
only one trajectory, called the escape path, arrives at the outer turning surface with zero velocity. Other trajectories, corresponding to different initial conditions, cannot reach the outer turning surface.

In this paper, we compare grid-based approaches to the LAPs with
the nudged elastic band (NEB) method that  was originally formulated in the context of molecular systems ~\cite{Garrett1983,Mills1994,Mills1995,Jonsson1998}. In NEB,  the minimum action path can be obtained iteratively by continuously shifting the pathway  to the nearest minimum action path \cite{Henkelman2000,Henkelman2000a,ASE}. A similar  approach is a growing string method \cite{Peters2004}. To provide more insights, we also employ the Euler-Lagrange (EL) method to compute the stationary action path. 

In addition to the LAP, another characteristic trajectory in the collective space is 
the minimum-energy path (MEP), sometimes referred to as the static path. 
The MEP can serve as a first, rough approximation to the LAP. It is obtained by computing the steepest descent line on the potential energy surface, which passes through the local minima and saddle points.  To find the MEP, a flooding, or
watershed,  algorithm has been applied \cite{Mamdouh1998,Moller2001,Iwamoto2002,Moller2004,Wang2019a}. The NEB approach can also be adopted to find the MEP and saddle points \cite{Asgeirsson2021}. (For a review of modern optimization methods for finding MEPs, see \cite{Sheppard2008,More2004}.)

This paper is organized as follows. In Sec.~\ref{sec:EDF} we 
define the basics concepts of the nuclear EDF approach as applied to nuclear fission.
Section~\ref{Sec: Methods algorithms} describes the path-optimization methods used.
The results of our calculations and an analysis of trends are presented in Sec.~\ref{sec:results}. 
 Finally, Sec.~\ref{sec:summary}  contains the conclusions of this work.

\section{Nuclear EDF approach to spontaneous fission}\label{sec:EDF}

\begin{figure*}[!htb]
        \includegraphics[width=0.8\linewidth]{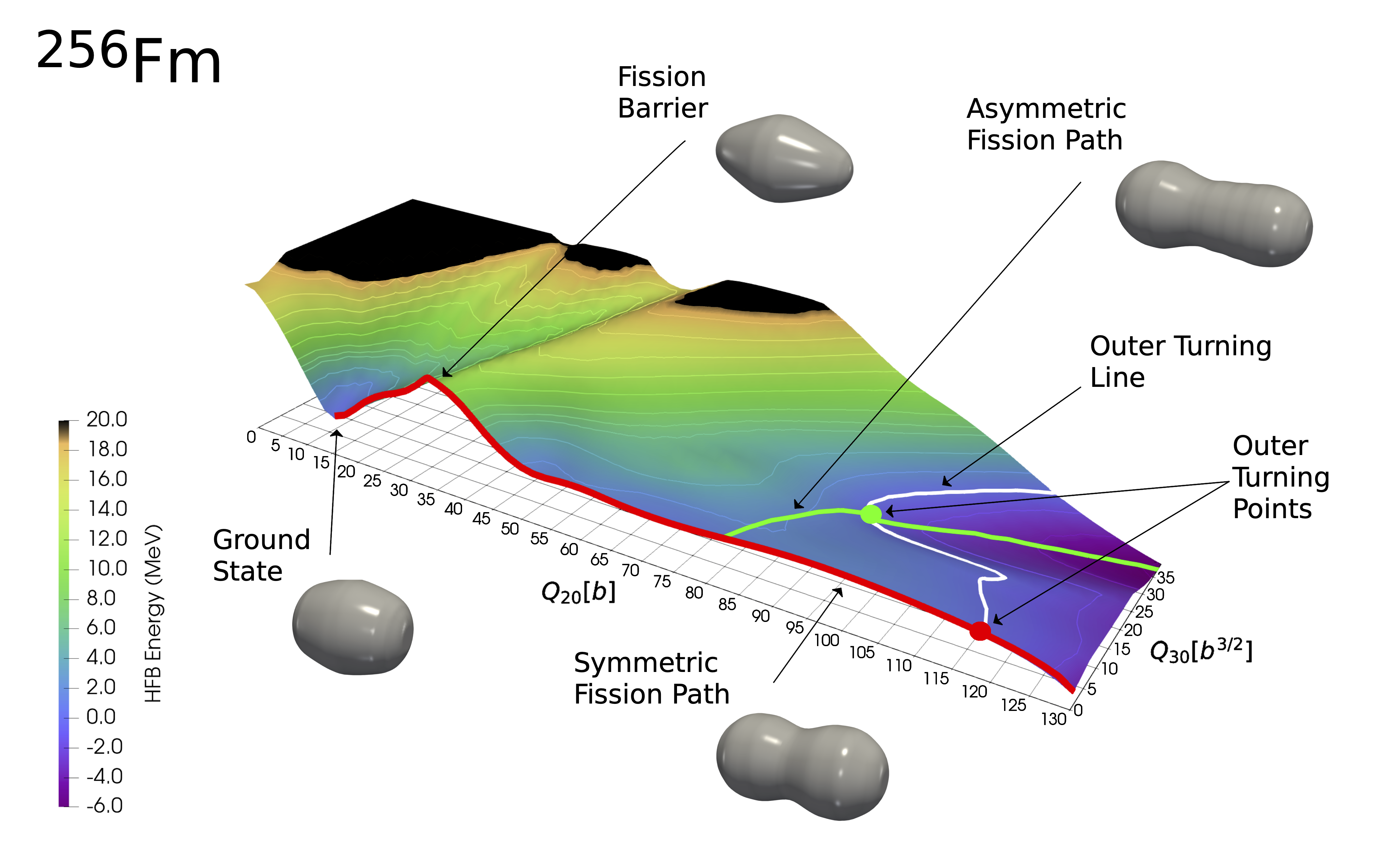}
        \caption{Potential energy surface of $^{256}$Fm  calculated with nuclear EDF method using the D1S parametrization of the Gogny interaction \cite{berger1984microscopic} in the space of two collective coordinates: $Q_{20}$ (elongation) and $Q_{30}$ (mass asymmetry). The static fission pathways are marked by solid lines: red (symmetric pathway) and green (asymmetric pathway). The outer turning line (OTL) is indicated, together with the outer turning points associated with the static pathways. For simplicity, we assume that the inner turning point corresponds to the ground-state configuration (i.e., $E_0=0$). The high-energy region that is practically not accessible  during collective motion is indicated in black. The intersections of fission pathways with outer turning points are indicated by dots; these are important for determining fission fragment yields \cite{Sadhukhan2020,Sadhukhan2022}. }
        \label{fig:PES}
\end{figure*}

The main ingredients for a theoretical determination of SF lifetimes are the collective potential
energy surface (PES)  and the inertia tensor. 
 To compute the PES, one solves the constrained Hartree-Fock-Bogoliubov (HFB) equations with the realistic energy density functional in the space of
collective coordinates $\vec{q} \equiv \{q_i\}$. 
These
are usually represented by  the  expectation values of the quadrupole moment operator $\hat{Q}_{20}$ (elongation), 
quadrupole moment operator $\hat{Q}_{22}$ (triaxiality), 
octupole moment operator $\hat{Q}_{30}$ (mass-asymmetry), and
the particle-number dispersion term $\lambda_{2\tau}(\hat{
N}^2_\tau - \langle \hat{ N}_\tau\rangle^2)$ ($\tau=n,p)$ that  controls dynamic pairing
correlations \cite{Vaquero2011,Vaquero2013,Sadhukhan2014}. 
In some cases one also considers the hexadecapole 
moment $Q_{40}$ (necking coordinate) \cite{Warda2012}. That is, in practical applications, we consider 2-5 collective coordinates which describe the collective motion of the system.
Figure~\ref{fig:PES} shows a representative PES of $^{256}$Fm in the space of 
$Q_{20}\equiv \langle \hat{Q}_{20}\rangle $  and $Q_{30}\equiv \langle \hat{Q}_{30}\rangle$.

The collective inertia (or mass) tensor $\mathcal{M}_{ij}(\vec{q})$ is obtained
from the self-consistent densities by employing the the adiabatic time-dependent HFB  approximation (ATDHFB) \cite{Baran2011,Giuliani2018b,Washiyama2021}.
In this study,
we use  the non-perturbative cranking approximation~\cite{Baran2011}:
\begin{equation}
\label{cranking-mass}
\mathcal{M}_{ij}(\vec{q})=\frac{\hbar^2}{2\dot{q}_i\dot{q}_j}\sum_{\alpha\beta}\frac{\left(F^{i*}_{\alpha\beta}F^{j}_{\alpha\beta}+
F^{i}_{\alpha\beta}F^{j*}_{\alpha\beta}\right)}{E_{\alpha}+E_{\beta}},
\end{equation}
where  $q_i$ is the collective coordinate, $\dot{q}_i$ represents the time derivative of
$q_i$, and $E_{\alpha}$ are
one-quasiparticle energies of HFB eigenstates $|\alpha\rangle$. The matrices $F^i$ are given by
\begin{equation}
\label{equation-F}
\frac{F^{i*}}{\dot{q}_i}=
 A^T\frac{\partial\kappa^*}{\partial q_i}A
             +A^T\frac{\partial\rho^*}{\partial q_i}B
-B^T\frac{\partial\rho}{\partial q_i}A
             - B^T\frac{\partial\kappa}{\partial q_i}B,
\end{equation}
where $A$ and $B$ are the  matrices of the  Bogoliubov transformation,
and $\rho$ and
$\kappa$ are particle and pairing density matrices, respectively, determined
in terms of $A$ and $B$. Derivatives of the density matrices with respect
to collective coordinates are calculated by employing the three-point Lagrange
formula. It is important to remark that  rapid variations in $\mathcal{M}_{ij}$ are expected in the regions of configuration changes (level crossings) due to
strong variations of density derivatives in (\ref{equation-F}) associated with structural rearrangements \cite{Ledergerber1973,Sadhukhan2013}.

Since SF is a quantum-mechanical tunneling process and the
fission barriers are usually both high and wide, the SF lifetime
is obtained semi-classically \cite{Brack1972} as $T_{1/2}=\ln2/(nP)$, where $n$ is the number of assaults on the fission barrier per unit time and $P$ is the penetration probability given by
\begin{equation}\label{penertration}
P=\left(1+\exp{[2S(L_{\rm min})]}\right)^{-1},
\end{equation} 
where $L_{\rm min}$ is the path that minimizes the fission action integral calculated along the one-dimensional trajectory
$L(s)$ in the multidimensional collective space:
\begin{equation}
\label{action-integral}
S(L)=\frac{1}{\hbar} \int_{s_{\rm in}}^{s_{\rm out}} {\cal S}(s)\,ds,
\end{equation}
where
\begin{equation}
\label{action-integrand}
{\cal S}(s) =\sqrt{2\mathcal{M}_{\text{eff}}(s)
\left(V_{\text{eff}}(s)-E_0\right)}
\end{equation}
with $V_{\text{eff}}(s)$ and
$\mathcal{M}_{\text{eff}}(s)$ being  the  effective potential energy and 
inertia along the fission path $L(s)$, respectively. 
$V_{\text{eff}}$ can be obtained by
subtracting the vibrational zero-point energy  from the total HFB energy. (In the examples considered in this paper we assume the zero-point energy to be zero.) The integration limits $s_{\rm in}$
and $s_{\rm out}$ correspond to the classical inner and outer turning points, respectively,  defined by $V_{\text{eff}}(s)=E_0$  on the two extremes of the fission path, see Fig.~\ref{fig:PES}. The collective ground state (g.s.) energy is $E_0$, and $ds$ is the element
of length along $L(s)$.  A one-dimensional path $L(s)$ can be defined in the multidimensional collective space by
specifying the collective variables $\vec{q}(s)$ as functions of path's length $s$.  The expression for $\mathcal{M}_{\text{eff}}$ is~\cite{Baran2005}:
\begin{equation}
\label{eff-mass}
\mathcal{M}_{\text{eff}}(s)=\sum_{ij}\mathcal{M}_{ij}(\vec{q})\frac{dq_{i}}{ds}\frac{dq_{j}}{ds}.
\end{equation}

\begin{figure}[htb!]
\label{Fig: Bifurcation}
        \includegraphics[width=\linewidth]{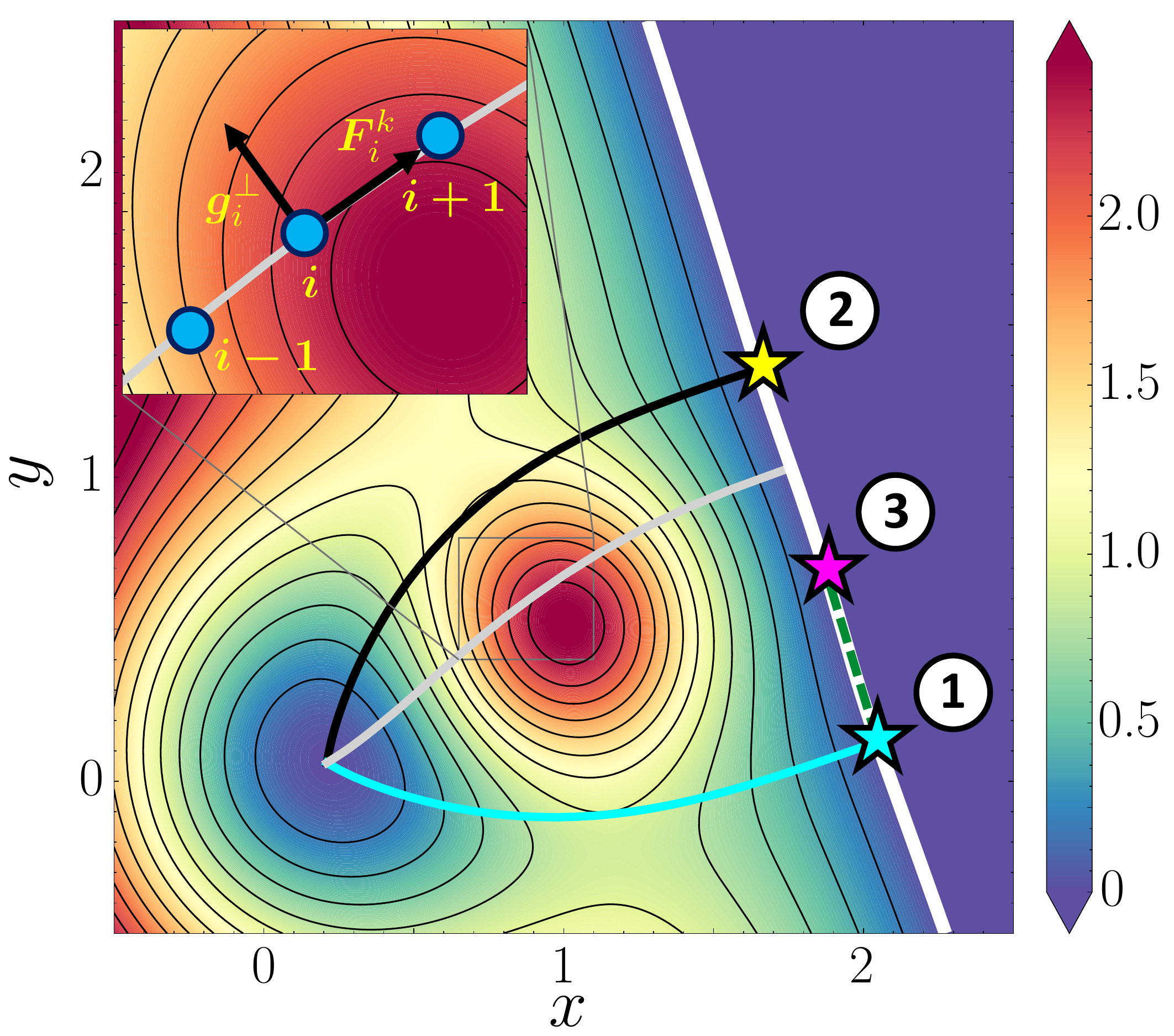}
    \caption{Illustration of two stationary action paths (representing competing fission modes) from the g.s. to the OTL (marked white) on the PES given by Eq.~(\ref{epsbimodal}). The cyan line shows the primary path (1). The secondary path (2) is indicated by the  black line. The corresponding exit points are marked by stars. The green dashed line connects  the exit point (1) with the point (3) on the OTL; the action along the dashed path is zero. The inset shows the spring force and the action force acting on the image $i$ on the NEB for the intermediate (not fully converged) grey path. For the video illustrating the NEB determination of both LAPs, see the supplemental material (SM) \cite{SM}.}
\end{figure}

The least-action path (LAP) $L_{\rm min}$ is obtained by minimizing the action integral  (\ref{action-integral}) with respect to all possible trajectories $L$  that connect the lines/surfaces of inner turning points $s_{\rm in}$ and outer turning points $s_{\rm out}$ \cite{Sadhukhan2013}. However, as discussed in Refs.~\cite{Schmid1986,Eckern1992} and this paper, only the pathways related to the exit points are stationary.
The MEP can instead be described as the union of steepest descent paths from the saddle
point(s) to the minima. The corresponding trajectory $\vec{q}(s)$ satisfies
\begin{align}
    \frac{d \vec{q}}{ds} \propto  \vec{\nabla} V\big(\vec{q}(s)\big)
    \label{MEP-def}
\end{align}
which characterizes a path of steepest descent on a surface $V(\vec{q})$ \cite{quapp1984analysis}. For the NEB, one finds the MEP by allowing the elements of the path to follow the gradient of the PES in their immediate vicinity.
We shall assume that the PES in the tunneling region is free from discontinuities associated with rapid configuration changes  \cite{Dubray2012,Zdeb2021,Lau2021}.
This assumption is usually valid because of non-vanishing pairing correlations inside the potential barrier. It is also to be noted that, as in any optimization/minimization approach, the stationary path determined numerically corresponds to a local action minimum, which is not guaranteed to be the global minimum. Moreover, there could be many stationary pathways representing different fission modes, see Fig.~\ref{Fig: Bifurcation}. To simplify notation, we assume in the following discussions that the stationary action path found by our algorithms is indeed the LAP.

 Since ${\cal S}(s)=0$  on the outer turning surface $V(\vec q) = E_{0}$, it follows that paths moving on the surface $V(\vec{q}) = 0$ do not contribute to the action. This is illustrated in Fig. \ref{Fig: Bifurcation} by the path connecting the g.s. and, for example, the purple star labeled (3). Such a path consists of the cyan curve -- the exit trajectory -- and the green dashed line, connecting (1) with (3) through the OTL, which results in the same action integral as the exit trajectory.

\section{Methods/algorithms} \label{Sec: Methods algorithms}

All path-optimization methods described in the following subsections, bar the EL method, have a reference implementation included in the python package, PyNEB \cite{PyNEB}.

\subsection{Nudged Elastic Band}\label{Sec: NEB}

The NEB method was originally formulated to provide a smooth transition of a molecular system on a potential energy surface from the reactant to the product state~\cite{Mills1994,Mills1995,Jonsson1998}.
Upon application of this variant of the NEB method, one obtains the MEP  as well as a series of ``images'' of the molecular system as it transitions along the path.
The NEB technique has been subsequently refined, with improved numerical stability~\cite{Henkelman2000a} and a more accurate determination of a saddle point~\cite{Henkelman2000} being two key advances towards a more widely applicable numerical approach for MEP determination.

To obtain the LAP, the procedure must be modified such that the images move towards the minimum of the action~\cite{Asgeirsson2018} which amounts to replacing the standard gradient of the PES with the gradient of the action
\begin{equation}
 \vec{g}_i= -\vec{\nabla}_i S
 \end{equation}
with respect to the image  $\vec q_i$.
With this prescription, the images will settle to the LAP in the collective space.

While the NEB method will, by design, drive the line of images towards either the MEP or LAP, the iterative scheme chosen greatly impacts the total number of iterations required before the solution converges.
In the early implementations, a simple velocity Verlet algorithm
\cite{Verlet1967} was used to adjust the position of the images step to step~\cite{Mills1994,Mills1995,Jonsson1998,Henkelman2000a,Henkelman2000}.
This approach is robust and relatively stable, though the convergence can be slow for flatter surfaces where the images are not pulled strongly to their optimal positions.
To aid this process, the Fast Inertial Relaxation Engine (FIRE) was proposed~\cite{Bitzek2006} to accelerate convergence without sacrificing stability. The method was subsequently updated~\cite{Guenole2020} to further improve performance.
Indeed, in our tests, the inertial algorithm regularly outperforms the velocity Verlet algorithm by 
an-order-of-magnitude reduction in iterations at the same convergence criteria.

With this, our implementation of the NEB approach is defined.
The algorithm itself is outlined in Algorithm~1 in  SM \cite{SM}. The force used in the optimization step for each image, $\vec{F}_i^{\rm opt}$, is constructed by adding the perpendicular component of the action gradient to the spring force $\vec{F}_i^{k}$ between the images, 
\begin{equation}
\vec{F}_i^{k} = k (|\vec{q}_{i+1} - \vec{q}_i| - |\vec{q}_i - \vec{q}_{i-1}|)\vec{\tau_i},
\end{equation}
where $k$ is a tunable parameter that controls the strength of the spring force and $\vec{\tau_i}$ is the unit vector tangent to the line of images from image $i-1$ to image $i+1$. 
The spring force on the endpoints is defined differently:
\begin{equation}
\vec{F}_1^k=k|\vec{q}_2-\vec{q}_1|,\quad 
    \vec{F}_N^k=k|\vec{q}_{N}-\vec{q}_{N-1}|.
\end{equation}
The total force acting on the interior images is then 
\begin{align}\label{Eq:ImagesForces}
    \vec{F}_{i}^{\rm opt} = \vec{F}^{k}_{i} + \vec{g}_{i}^{\perp}.
\end{align}

The NEB approach is illustrated in Fig.~\ref{Fig: Bifurcation} for the case of bimodal tunneling from  the g.s. minimum to the OTL on an analytic  PES defined by: 
\begin{align}\label{epsbimodal}
    V(\vec{q}) &= 3.17 + 2e^{-5\big((x-1)^{2} + (y-\frac{1}{2})^{2}\big)} -3 e^{-(x^{2} +y^{2})} \nonumber \\&-\frac{1}{2}(3x +y),
\end{align}
where $\vec{q} = (x,y)$.
The inset shows the forces on the images of the NEB grey path, which has not converged yet to the black path. The spring force $\boldsymbol{F}^k_i$ keeps the images from drifting too much from each other, while the perpendicular part of the action gradient $\boldsymbol{g}_i^\perp$ pushes them towards the nearest stationary action path. This example shows that the NEB algorithm, depending on the initial locations of the images, will converge to a local stationary path, not necessarily the least action path.

For the endpoint, $i=N$, one can choose to either fix the position of the image or to allow the image to move towards the outer turning surface.
In the second case, a harmonic restraint term is added to the spring force to construct $\vec{F}_{N}^{\rm opt}$,
\begin{equation}
\label{eqn:endpoint}
    \Vec{F}_{N}^{\rm opt} = \Vec{F}_{N}^{k} - \left[\Vec{F}_{N}^{k} \cdot \vec{f}(\vec{q}_{N}) - \eta (V(\vec{q}) - E)\right]\vec{f}(\vec{q}_{N}),
\end{equation}
where $\vec{f} = -\vec\nabla V / |\vec\nabla V|$  and $\eta$ determines the strength of the harmonic restraint term~\cite{Asgeirsson2018}.
This force pulls the endpoint $i=N$ very quickly to the outer turning surface and helps find the optimal outer turning point. 

The default iteration scheme used in our implementation is the inertial algorithm mentioned above, though a standard Verlet minimizer is also included in the PyNEB python package \cite{PyNEB}.
The structure of the NEB solver is modular and allows for the simple replacement of components like the minimizer, allowing for easy checks on the convergence and parameters that describe the iterative scheme.


\subsection{Grid-Based Methods}\label{Section: grid-methods}
Some traditional methods to compute the LAP begin by computing the PES and the collective inertia on a grid of collective coordinates. The calculation of the LAP is then reduced to finding the path through the grid points that minimizes a discrete approximation of the action. Two methods that we have benchmarked are the dynamic programming method (DPM) \cite{Baran1981}, and Dijkstra's algorithm (DA) \cite{Dijkstra}. Here, both will be described for two-dimensional (2D) grid, with points labelled by $\vec{q}_{ij}=(x_i,y_j)$ ($i=1,\ldots,N$, $j=1,\ldots,M$). Both methods can be straightforwardly extended to a higher-dimensional grid.

Dynamic programming is a general mathematical technique for solving multi-decision problems by breaking the problem down into simpler overlapping sub-problems.  It was first adapted to the action integral  minimization in Ref.~\cite{Baran1981} and used in \cite{Sadhukhan2013} to determine the LAP. This adaptation is what we refer to as the DPM.

\begin{figure}[htb!]
\label{grid_method}
        \includegraphics[width=0.48\textwidth]{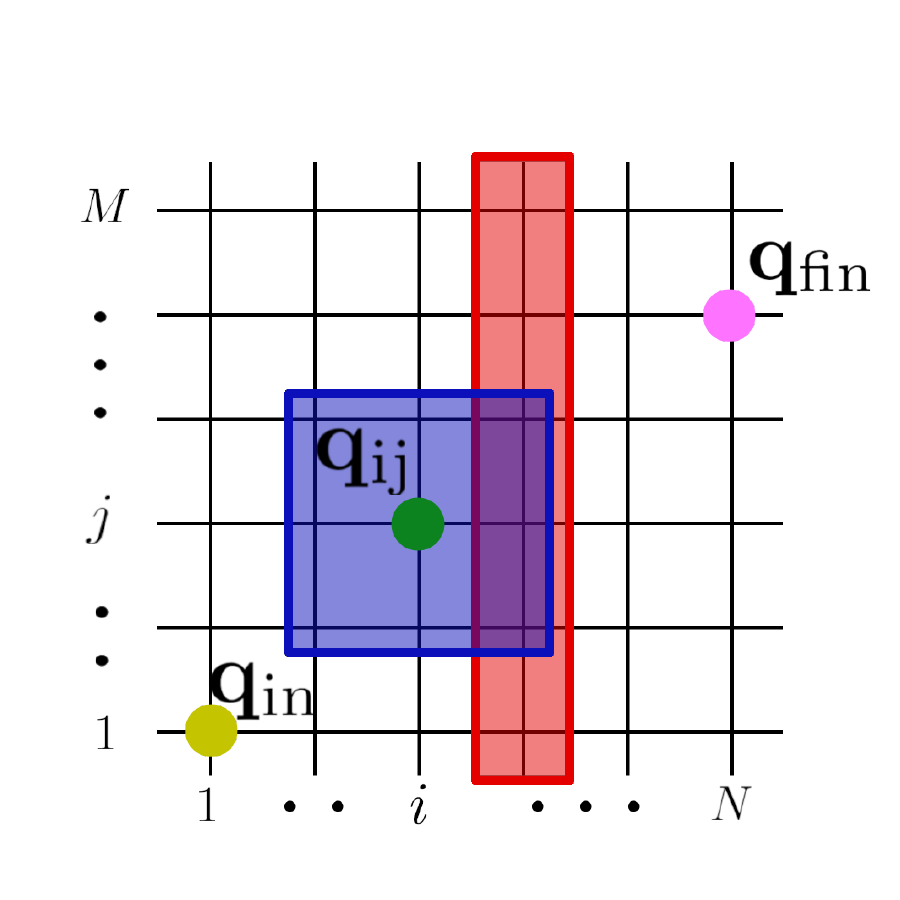}
    \caption{Different types of paths that can be found in the different grid-based methods. The single node $\vec{q}_{ij}$ can reach the red (blue) regions in DPM (Dijkstra's algorithm). The initial and final points are  marked.}\label{Fig: grid_methods}
\end{figure}

The DPM approximates the LAP between an initial point, $\vec{q}_\textrm{in}$, and a final point, $\vec{q}_\textrm{fin}$. This method finds paths that traverse diagonally from a given cell: from cell $\vec{q}_{ij}$, only cells $\vec{q}_{i+1,j}$ can be reached, for $j=1,\ldots,M$. The allowed cells are highlighted in red in Fig.~\ref{grid_method}. The LAP from $\vec{q}_\textrm{in}$ to $\vec{q}_\textrm{fin}$ is constructed iteratively as follows: for a cell $\vec{q}_{ij}$, there are $M$ possible paths, each passing through a cell at $x_{i-1}$. The LAP from $\vec{q}_\textrm{in}$ to $\vec{q}_{ij}$ is selected and stored in memory. This is repeated for every cell with $x=x_i$, for a total of $M$ possible paths. Once $\vec{q}_\textrm{fin}$ is reached, there are only $MN$ paths (out of a total of $M^N$ paths), and the LAP is selected from these. The DPM algorithm is detailed in Algorithm 2 in  SM \cite{SM}.

Dijkstra's method \cite{Dijkstra} is similar to DPM, in that it breaks down the large optimization problem into a set of smaller problems. Given a cell $\vec{q}_{ij}$, the action to every neighbor $\vec{q}_{i'j'}$ is calculated as if the path to $\vec{q}_{i'j'}$ passes through $\vec{q}_{ij}$. If this action integral is smaller than that along the current path to $\vec{q}_{i'j'}$, $\vec{q}_{i'j'}$ is said to come from $\vec{q}_{ij}$. This is repeated, starting from $\vec{q}_\textrm{in}$, until $\vec{q}_\textrm{fin}$ is reached. Figure \ref{grid_method} shows the nearest-neighbors of $\vec{q}_{ij}$ (the cell marked in green) in a blue square. Dijkstra's algorithm is described in Algorithm~3 in  the SM \cite{SM}.

Dijkstra's algorithm can find paths that pass through multiple cells with the same $x_i$ value, or even paths that backtrack. DPM cannot find such paths. However, DPM can find paths that jump from $\vec{q}_{ij}$ to $\vec{q}_{i+1,j'}$, for any $j'$, while Dijkstra's algorithm is limited to $j'=j-1,j,j+1$ (see Fig.~\ref{grid_method}). For fission calculations, one frequently takes the $x$ coordinate as the quadrupole moment $Q_{20}$, and fission can be viewed as collective motion in which $Q_{20}$ continuously increases towards scission. So, the paths that Dijkstra's algorithm can find, that DPM cannot, are rather unlikely.  In general DPM tends to find paths with a smaller action than Dijkstra's algorithm, see Sec.~\ref{sec:results}.

\subsection{Euler Lagrange Equations}

In order to find the LAP for the functional~\eqref{action-integral} using the EL  equations \cite{weinstock1974calculus}, we first parametrize the trajectory $\vec q$ by a time variable $t$,
i.e., $\vec{q}=\vec q(t)$ with $t \in [0,t_f]$. This is done in order to explicitly account for the arclength $ds = (\sum dq_i^2)^{1/2}$. In terms of $t$, the action integral~\eqref{action-integral}  reads:
\begin{equation}\label{Eq: ELE Action}
\begin{split}
    &S(L) = \\
    &\int_{0}^{t_f} \sqrt{2\big(V_\text{eff}[\vec q(t)]-E_0 \big)}\Big(\sum_{ij}^n\mathcal{M}_{ij}[\vec q(t)]\dot{q}_i\dot{q}_j\Big)^{1/2}dt \\
    &= \int_{0}^{t_f} \mathcal{L}(\vec q,\dot{\vec q})dt,
\end{split}
\end{equation}
where  $\dot{q}_i\equiv dq_i/dt$, and $\mathcal{L}$ is the corresponding Lagrangian. The associated EL equation can be written as:
\begin{equation}\label{Eq: ELE}
    \frac{\partial \mathcal{L}}{\partial q_i} = \frac{d}{dt} \Big( \frac{\partial \mathcal{L}}{\partial \dot{q}_i}\Big),
\end{equation}
with the boundary conditions:  $\vec q(t=0)=\vec q_\text{in}$ (the initial location) and $\vec q(t=t_f)=\vec q_\text{\textrm{fin}}$ (the final location). 

In order to numerically solve Eq.~\eqref{Eq: ELE} we use the shooting method \cite{Shooting}. That is, we start at the initial position $\vec q_\text{in}$ and vary the direction and orientation of the initial ``velocity'' $\dot{\vec q}(t=0)$. We use a numerical differential equation solver to propagate the solution until we find an initial condition that satisfies $\vec q(t_f)=\vec q_{\textrm{fin}}$. Finding such initial conditions can present some challenges, which we discuss in the SM \cite{SM}.

The EL approach is equivalent to what is done in Ref.~\cite{Schmid1986} where the eikonal equation is solved by the method of characteristics. Each different trajectory  obtained by varying $\dot{\vec q}(t=0)$ corresponds to one of the characteristics of the leading order (cf. Eqs.(2.8) and (4.3) of \cite{Schmid1986}).
It is worth noting that if the imaginary part of the phase of the wave function $W(\boldsymbol{q})$  is  negligible, as is the case of the motion in the deep subbarrier region, then the eikonal equation for $W$ is a valid approximation \cite{Kapur1937}.  The trajectories  corresponding to   the stationary functional~\eqref{action-integral} are equivalent to the solutions of the eikonal equation for $W$ (see Eqs.~(11) and (13) of \cite{Kapur1937}). A connection between the eikonal equation, the dynamic programming approach, and a variational principle in the context of geometrical optics is discussed in Ref.~\cite{lakshminarayanan1997dynamic}.

\section{Results}\label{sec:results}

\subsection{Analytic surfaces: Illustrative examples}\label{Sec: Analytic Surfaces}

We  benchmark the performance of the NEB method by comparing the LAP found using NEB (denoted as NEB-LAP) to the paths found using the DPM, DA, and EL approaches for  analytic surfaces defined in terms of the position vector $\vec{q}=(x,y)$.
Throughout this section, we assume a constant inertia ${\cal M}_{ij}=\delta_{ij}$.
Within the NEB framework, the action functional \eqref{action-integral} can develop some noise as the NEB algorithm  approaches the final action. This noise is a function of the NEB hyperparameters and the optimization method used.  All surfaces discussed in this section are released as example cases with PyNEB
\cite{PyNEB}.

In the analytic cases, the NEB is initialized by fixing an initial and final points $\vec{q}_{\textrm{in}}$ and $\vec{q}_{\textrm{fin}}$, respectively, and defining a linear trajectory connecting them.   The NEB algorithm  is then iterated until convergence is reached. Grid-based methods use a grid spacing of $\Delta x = 0.1$ along the x-axis and $\Delta y = 0.005$ along the y-axis for all analytic surfaces. Details of the numerical methods used for solving the EL equations for all surfaces are discussed in  SM \cite{SM}. The action values for each surface considered are included in Table \ref{analytic-table}.  Action integrals in Table~\ref{analytic-table} are evaluated using linearly interpolated trajectories over 500 uniformly-distributed points. 

We compute both LAP and MEP in the NEB framework. Since the MEP is a solution of  Eq.~(\ref{MEP-def}), images along the path  converge to critical points on the surface depending on the position of the boundary images  at $\vec{q}_{\textrm{in}}$ and $\vec{q}_{\textrm{fin}}$. Critical points on the surface $V(\vec{q})$ contained in the MEP can be extracted by calculating $\vec{\nabla}V$ along the path and are classified by computing the eigenvalues of the Hessian at those points. 
\begin{table}[htb]
    \caption{Action integrals for the 6-Camel-Back  (CB-S and CB-A) and M{\"u}ller-Brown (MB) surfaces. The integrals have been calculated using a linear spline interpolation evaluated at 500 points along each trajectory.}
    \label{analytic-table}
    \begin{ruledtabular}
        \begin{tabular}{cccccc}
            &
            \textrm{NEB-MEP}&
            \textrm{NEB-LAP}&
            \textrm{DPM}&
            \textrm{EL}&
            \textrm{DA}\\
            \colrule
            CB-S & 5.522 & 5.518 &  5.524 & 5.536 & 5.563\\
            CB-A & 6.793 & 6.404 &  6.405 &  6.407 & 6.886\\
            MB & 28.491 & 22.875 & 22.909 & 22.871 & 23.427\\
        \end{tabular}
    \end{ruledtabular}
\end{table}

\begin{figure}[t!]
    \includegraphics[width=0.5\textwidth]{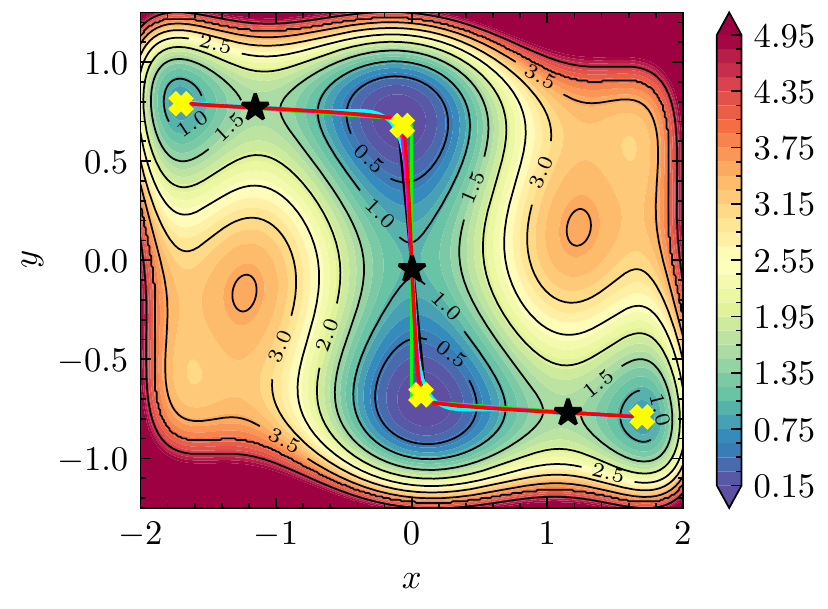}
    \caption{The symmetric camel-back PES $V_{CB-S}(\vec{q})$ normalized to  its global minimum together with the calculated NEB-MEP (red), NEB-LAP (magenta), DPM (black), EL (cyan), and DA (lime) trajectories. Black stars indicate  saddle points and yellow crosses mark local minima.}
    \label{fig-camel-symmetric}
\end{figure}

First, we consider the symmetric 6-Camel Back potential (CB-S) \cite{More2004} defined as
\begin{equation}
    V_{\rm CB-S}(\vec{q}) = \big(4 - 2.1 x^{2} + \frac{1}{3} x^{4} \big) x^{2} + x y + 4(y^{2} - 1) y^{2}
    \label{eq:cbs}
\end{equation} 
In this example, we seek the LAP connecting the local minimum located at $\vec{q}_{\textrm{in}} = (1.70, -0.79)$ to the local minimum located at $\vec{q}_{\textrm{fin}} = (-1.70, 0.79)$. Figure \ref{fig-camel-symmetric} shows the CB-S PES 
normalized to zero at its global minimum  together with the calculated NEB-MEP, NEB-LAP, DPM, EL, and DA trajectories. The action integrals along these 
trajectories are listed in Table \ref{analytic-table}.
\begin{figure}[t!]
    \includegraphics[width=0.5\textwidth]{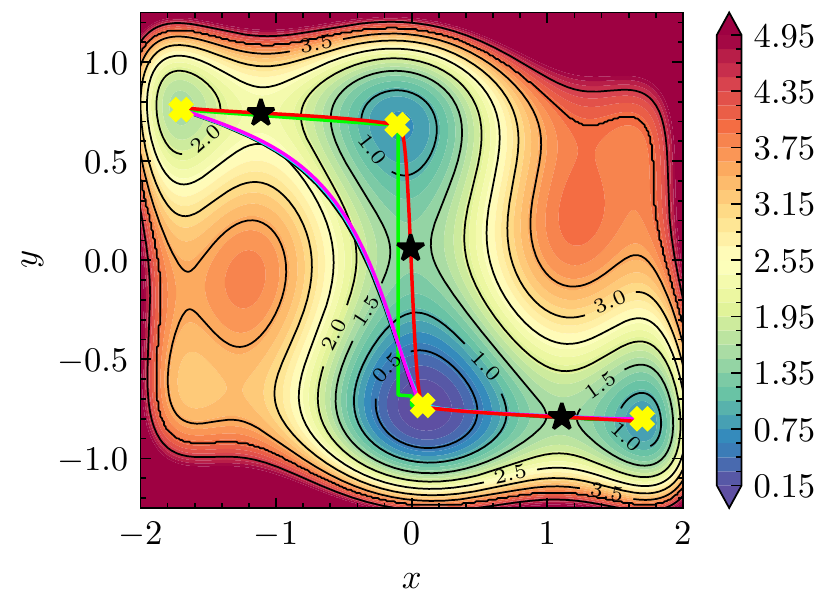}
    \caption{Similar as in Fig.~\ref{fig-camel-symmetric} but for the asymmetric camel-back  surface $V_{\textrm{CB-A}}(\vec{q})$.
    For the video illustrating the NEB determination of both LAP and MEP, see the SM \cite{SM}.}
    \label{fig-camel-asymmetric}
\end{figure}
The MEP and the LAPs computed by using the NEB, EL, DPM, and DA methods are very similar. However,  the DA trajectory slightly deviates from the other ones. This is because DA is more  constrained by the grid spacing than DPM: regardless of the grid spacing, DA can only consider its immediate neighbors, while DPM does not have this constraint (see Fig.~\ref{Fig: grid_methods} and \ref{Section: grid-methods}).

As indicated by Fig. \ref{fig-camel-symmetric}, the final action values for the LAP obtained by the NEB, DPM, and EL methods agree well with the MEP. However, the MEP and LAP are not necessarily equivalent in general; the MEP can be viewed as an approximation of the LAP. A detailed discussion on the conditions for the MEP to be an LAP is contained in  the SM \cite{SM}. To see the MEP limitations, we consider an asymmetric variant of the Camel-Back potential  (CB-A)
\begin{align}
    V_{\rm CB-A}(\vec{q}) = V_{\rm CB-S}(\vec{q}) + \frac{1}{2}y
\end{align}
where the end points of the local minima are $\vec{q}_{\textrm{in}} = (1.70, -0.8)$ and $\vec{q}_{\textrm{fin}} = (-1.70, 0.76)$. Figure~\ref{fig-camel-asymmetric} shows the MEP  trajectory which is markedly different from the LAP solutions and corresponds  to an appreciably larger action integral.  Still, the MEP can be used  for finding critical points (minima and saddles)  on the surface.

\begin{figure}[t!]
    \includegraphics[width=\linewidth]{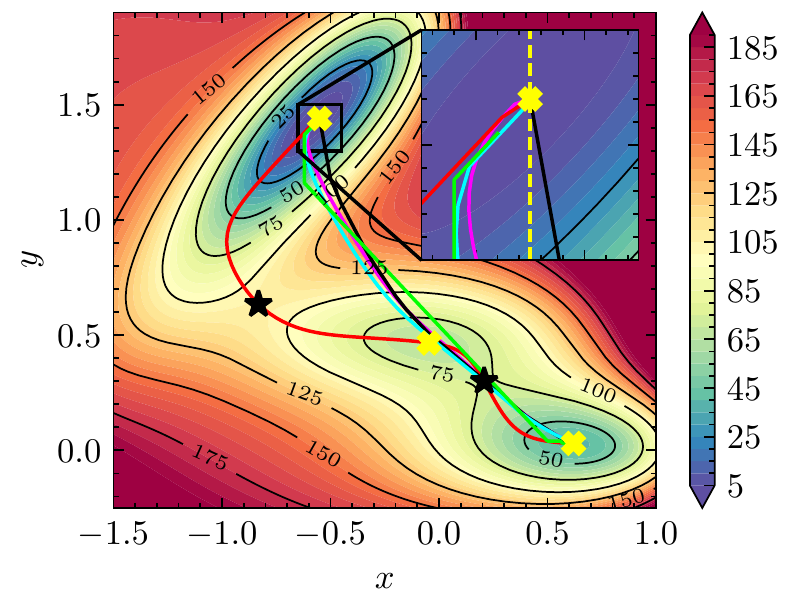}
    \caption{Similar as in Fig.~\ref{fig-camel-symmetric} but  for the shifted M{\"u}ller-Brown surface. The inset shows the LAP pathways close to the initial point $\vec{q}_\textrm{in}$. The yellow dashed line shows the vertical. As can be seen, all paths except for the DPM curve start by moving to the left of the vertical.}
    \label{fig-MB}
\end{figure}

The M{\"u}ller-Brown potential
 is a canonical example of a PES used in theoretical chemistry  \cite{muller1979location,Koistinen2017,Asgeirsson2018}.
The M{\"u}ller-Brown surface shown in Fig.~\ref{fig-MB} is defined as
\begin{equation}
    V_{\rm MB}(\vec{q}) = \sum_{i=1}^{4} A_{i}e^{a_{i}(x-x_{0_{i}})^{2} + b_{i}(x-x_{0_{i}})(y-y_{0_{i}}) + c_{i}(y-y_{0_{i}})^{2}},
\end{equation}
where we use the same set of parameters as in Ref.~\cite{muller1979location}, namely:  $\vec{A} = (-200, -100, -170, 15)$, $\vec{a}=(-1,-1,-6.5,0.7)$, $\vec{b}=(0,0,11,0.6)$, $\vec{c}=(-10,-10,-6.5,0.7)$, $\vec{x}_{0}=(1,0,-0.5,-1)$, and $\vec{y}_{0}=(0,0.5,1.5,1)$.
The MEP follows the bent trajectory that goes through the critical points: two saddle points and one local minimum. This trajectory markedly differs from the LAPs,
which are in a rough agreement. The MB surface highlights a problem with the DPM. As mentioned in Sec.~\ref{Sec: Methods algorithms}, the DPM can only search a single direction of each coordinate axis of the domain. In the case of the Muller-Brown surface, the DPM cannot search for trajectories bending back in the negative-$x$ direction. As seen in the inset of Fig.~\ref{fig-camel-symmetric},  the NEB, EL, and DA methods start their trajectories moving backwards in $x$ from the initial point $\vec{q}_\textrm{in}$. The DPM path, on the other hand, always moves in the positive-$x$ direction. Consequently, the action integral along the DPM path is slightly larger than in the other methods. 

\subsection{Realistic calculations}\label{Sec: Realistic}

To illustrate the performance of the NEB method and other approaches 
to the LAP in realistic cases, we carried out nuclear EDF calculations for  $^{232}$U in two collective coordinates and $^{240}$Pu in three collective coordinates. In the particle-hole channel we used the Skyrme functional SkM$^{*}$ \cite{BARTEL198279}, which is often employed in fission studies. The particle-particle interaction was approximated by the mixed density-dependent pairing force~\cite{DobaczewskiPairing}.

In the case of  $^{232}$U, we considered two collective coordinates
$\vec{q}\equiv(Q_{20},Q_{30})$ and for $^{240}$Pu we took  three collective coordinates $\vec{q}\equiv(Q_{20},Q_{30},\lambda_2)$.
The axial quadrupole and octupole  moment operators are  defined as in  Ref.~\cite{DobaczewskiHFODD}:
\begin{align}
    \hat{Q}_{\lambda 0}(r,\theta)        = \mathcal{N}_{\lambda} \sqrt{\frac{2\lambda+1}{4\pi}}                            r^{\lambda} P_{\lambda} (\cos\theta)
\end{align}
where $P_{\lambda}$ is the Legendre polynomial, $\mathcal{N}_{2}=\sqrt{\frac{16\pi}{5}}$, and $\mathcal{N}_{3}=1$.  The collective coordinate $\lambda_2=\lambda_{2n}+ \lambda_{2p}$ defined in Sec.~\ref{sec:EDF}  represents the dynamic pairing fluctuations. The value of  $\lambda_{2\tau}$= 0 corresponds to static HFB pairing.

As in Ref.\,~\cite{Sadhukhan2014}, to render collective coordinates dimensionless, we  use dimensionless coordinates ${x_i}$ defined as
\begin{eqnarray}\label{eqn:scale_param}
    x_{i}&=&\frac{q_{i}}{\delta{q_{i}}},
\end{eqnarray} 
where $\delta{q_{i}}$ are the scale parameters used in determining numerical derivatives of density matrices in Eq.~(\ref{equation-F}).
Here we took  $\delta{Q_{20}} = 1$\,b, $\delta{Q_{30}} = 1$\,b$^{3/2}$ and $\delta{\lambda_2}= 0.01$\,MeV.

\subsubsection{Two dimensional case: SF of $^{232}$U}

The PES was computed by solving the HFB equations using the parallel axial solver HFBTHO(v3.00)\cite{PEREZ2017363}.  The large stretched harmonic oscillator basis of $N= 25$ major shells was used to guarantee good convergence. We adopted a  $458\times501$ grid with $0\le Q_{20} \le{457}$\,b and $0\le  Q_{30}\le{50}$\,b$^{3/2}$. To apply the NEB method, which involves local gradient calculations at arbitrary values $\vec q$,  we interpolate the PES and the inertia tensor on the mesh. Because the grid is two dimensional, a cubic spline interpolator suffices. Close to the $Q_{30}=0$ axis, we take into account the mirror symmetry of the PES  by setting $V(-Q_{30})=V(Q_{30})$. Finally, since  NEB updates  occasionally push an image outside of the computed PES mesh, we extended the PES to grow exponentially with the distance outside the mesh, to smoothly push images back into the evaluated region.

\begin{figure}[htb!]
                \includegraphics[width=1.0\columnwidth]{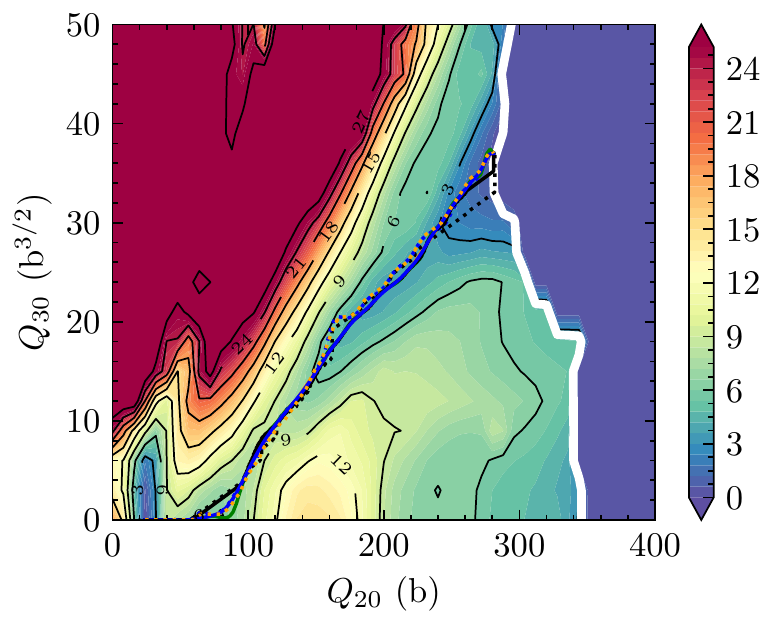}
        \caption{The PES of $^{232}$U in the $(Q_{20},Q_{30})$ plane calculated with SkM$^*$. Solid lines mark the LAPs and MEP obtained  with the constant  inertia tensor; dotted lines correspond to the non-perturbative inertia tensor. 
        The OTL is  shown in white.
        The blue, orange, purple and black curves represent the LAPs calculated using the NEB, DPM, EL, and DA methods, respectively. The green curve is the MEP, which was also calculated using NEB.}\label{Fig: U232 Paths}
\end{figure}

Figure~\ref{Fig: U232 Paths} shows the two-dimensional PES of $^{232}$U. The least action fission pathway which goes from the g.s. at
$\vec{q}_\textrm{in}$=(24\,b, 0) to the  exit point $\vec{q}_\textrm{fin}$=(281\,b, 37\,b$^{3/2}$) is calculated using the  methods explained in Sec.~\ref{Sec: Methods algorithms}.  To select the endpoint $\vec{q}_\textrm{fin}$, we compute the LAP using DPM for all points on the OTL, and select the point with the lowest action integral. This point is then used as the exit point for the other methods. While NEB does not require a fixed endpoint in general, we fix the endpoint here in order to facilitate inter-method comparison. The MEP path is calculated using the NEB method.

The action integral computed with different methods is shown in Table \ref{table:nuclei_acts}. When computing the action, we interpolate the paths using a linear spline interpolator, and the action integral  is computed using 500 evaluations along the path. This reduces the differences in the action that may arise from using a different number of points along the path (for instance, NEB gives a similar path to DPM using as few as 30 images). For all paths, we compute the action using the inertia tensor evaluated along the path. As can be seen, the action values computed for $^{232}$U using different methods agree well, with DA being the worst performer. As seen in Fig.~\ref{Fig: U232 Paths} and Table\,\ref{table:nuclei_acts} the MEP is very close to the LAP. This is because the static fission pathway (i.e., MEP) is fairly straight and the fission valley is well delineated.  Note that perfect agreement is not expected, and in fact was not observed for the analytic surfaces, either. This is due in part to the different approximations used in each method --- for DPM and DA, this is the grid spacing; for NEB, this is the number of images and approximate treatment of derivatives; and for EL, this is a variety of simplifications described in Sec. 3 in the SM \cite{SM}. Additional variation in the quality of the interpolator further hampers agreement beyond what is listed.

\begin{table}[htb]
\caption{Action integrals for $^{232}$U computed with different methods.  The paths computed using the constant and non-perturbative inertia tensor are labelled as  ``con.'' and ``n.-p.'', respectively.}
\label{table:nuclei_acts}
    \begin{ruledtabular}
        \begin{tabular}{llccccc}
            & &
            \textrm{NEB-MEP}&
            \textrm{NEB-LAP}&
            \textrm{DPM}&
            \textrm{EL}&\textrm{DA}\\
            \colrule
            \multirow{2}{*}{${}^{232}$U} & con.     & 174.5         &     174.2 & 174.2 & 174.9 & 175.8 \\
 & n.-p. & -         &     173.6 & 173.3 & 175.0 & 178.5 \\\\[-5pt]
                        \multirow{2}{*}{${}^{240}$Pu} & con.     & 19.09     &     18.98 & 19.21 & 19.01 & 22.85 \\
 &  n.-p. & -         &     16.54 & 16.47 & 18.18 & 30.50 \\
        \end{tabular}
  
  \end{ruledtabular}
\end{table}

\subsubsection{Three dimensional case: SF of $^{240}$Pu}

The SF of $^{240}$Pu in several collective coordinates was studied in Ref.~\cite{Sadhukhan2014} where the details pertaining to the computation, grid size, etc., can be found. Between the g.s. minimum and the fission isomer (FI),
the fission pathway is affected by triaxial degrees of freedom. Between the FI and the outer turning surface (OTS), however, the predicted fission trajectory is axial. In this paper, we consider the fission of the  FI of $^{240}$Pu so the OTS corresponds to the FI energy.

\begin{figure}[htb!]
    \includegraphics[width=1.0\columnwidth]{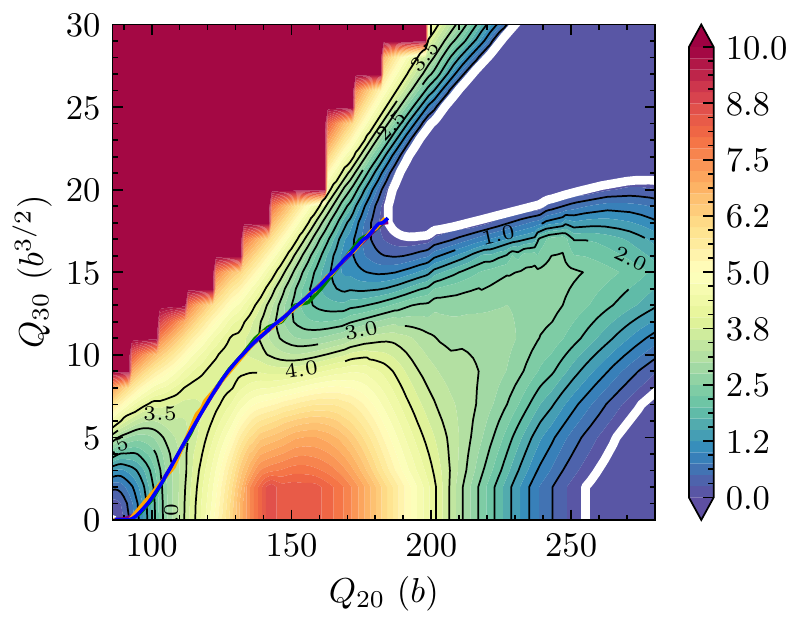}
    \caption{The PES for $^{240}$Pu in the space of collective coordinates $Q_{20}$, $Q_{30}$ with $\lambda_2 = 0$. Only the region beyond the fission isomer is shown. The energy is normalized to the energy of the fission isomer.  The OTL is  shown in white.
    The MEP (green) practically coincides with the LAPs calculated  with the constant inertia using  the NEB (blue), DPM (orange), and EL (purple) methods. }\label{Fig: Pu240 2d Paths}
\end{figure}

For three-dimensional tunneling,  the system of equations that must be solved to construct a global spline interpolator is too large for practical applications. Instead, we use piecewise linear interpolation. The PES at  $\lambda_2=0$   for $^{240}$Pu shown in Fig.~\ref{Fig: Pu240 2d Paths} varies very smoothly in the barrier region where the potential energy is larger than the energy of the FI, and so this interpolation scheme is reasonable. 

\begin{figure}[htb!]
        \includegraphics[width=1.0\columnwidth]{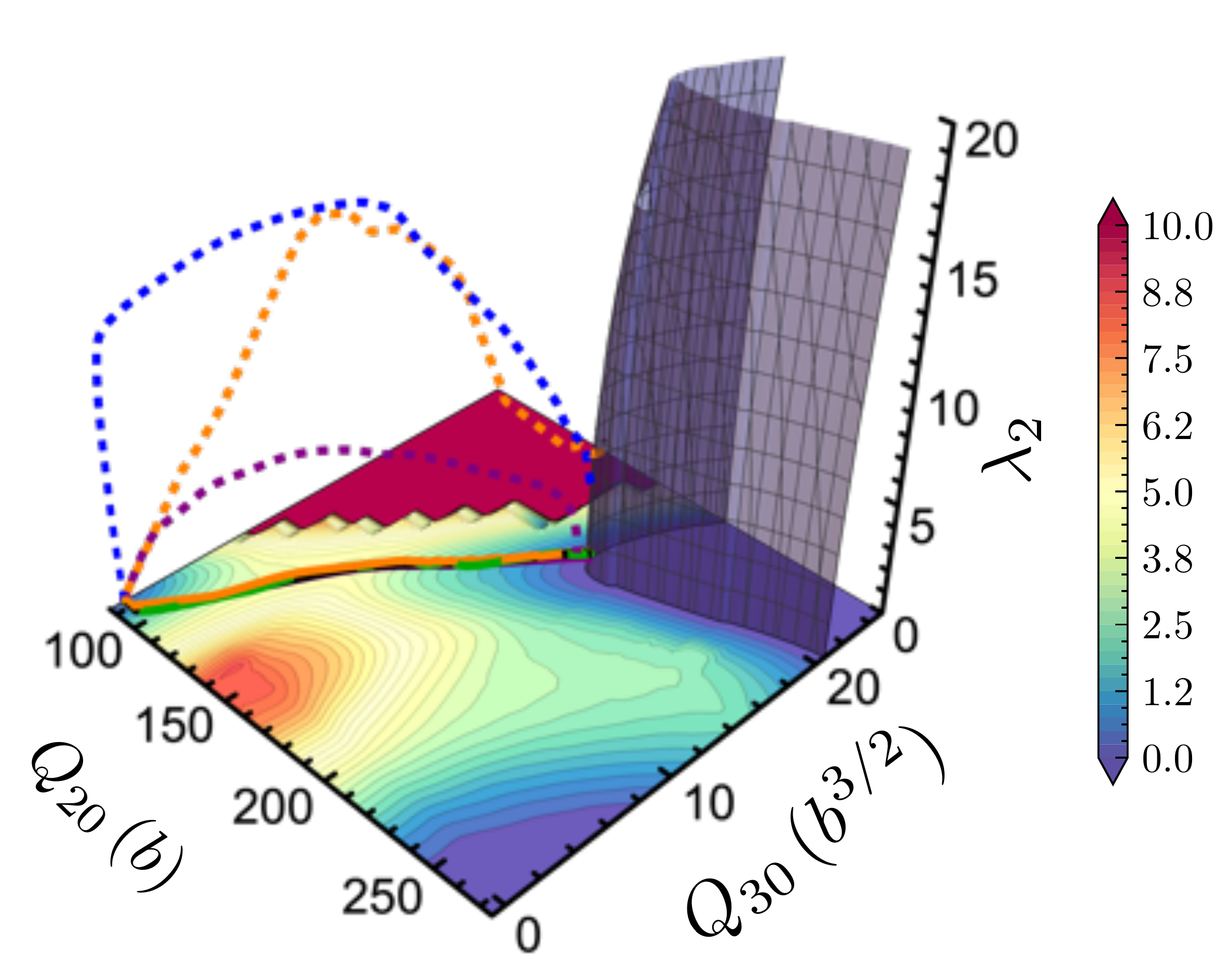}
        \caption{The PES for $^{240}$Pu in the collective coordinates $Q_{20}$, $Q_{30}$ and $\lambda_2$. The 2D     cross section at  $\lambda_2=0$ shown in Fig.~\ref{Fig: Pu240 2d Paths} is indicated. The blue, orange, and purple curves are the LAP, calculated using the NEB, DPM, and EL methods, respectively. The non-perturbative inertia tensor was used for the dashed curves. The OTS  is indicated by the dark blue contour surface.
        } 
        \label{Fig: Pu240 Paths}
\end{figure}

Figure~\ref{Fig: Pu240 Paths}  shows the LAPs for $^{240}$Pu
computed with the NEB, DPM, and EL methods in three dimensions (3D). The pathways begin at the FI minimum at  $\vec{q}_\textrm{in}=(Q_{20}^{0}=87\,\text{b}, Q_{30}^{0}=0\,\textrm{b}^{3/2}, \lambda_{2}=0.0)$ and the exit point was chosen for DP in the same way as the $^{232}$U results before.
The NEB endpoint in this case was allowed to vary according to Eq.~\ref{eqn:endpoint}, better representing standard procedure for production runs.
The exit points $\vec{q}_\textrm{fin}$ predicted by NEB ($185.1\,\text{b}, 18.4\,\text{b}^{3/2}, 3.3\,\text{MeV}$), DPM ($184.0\,\text{b}, 18.6\,\text{b}^{3/2}, 4.8\,\text{MeV}$) and EL ($179.8\,\text{b}, 17.7\,\text{b}^{3/2}, 0.0$) then differ.
When the collective mass is held constant, all methods find very similar paths in the $\lambda_{2}=0.0$ plane, which are also shown in Fig.~\ref{Fig: Pu240 2d Paths}.
The paths vary more when the non-perturbative inertia tensor is used, with the main difference between the NEB and DPM paths appearing in the region close to the FI minimum; beyond the saddle point, both paths are similar.

As seen in Table \ref{table:nuclei_acts}, the  NEB and DPM are in a good agreement. In general, one would expect a better performance from NEB as this method is not constrained to a grid (this is true in the case of the analytic surfaces discussed in Sec. \ref{Sec: Analytic Surfaces}). However, in rare cases, the DPM produces a slightly lower action than the NEB.
In such cases the NEB converges to an even lower action if is initialized with the DPM result.
This suggests that for tunneling in more than 2D, a combination of NEB and DPM might be beneficial.

\section{Conclusions}\label{sec:summary} 

Finding the path that minimizes the action integral can be extremely challenging since it involves searching over the space of all continuous paths that fulfill the boundary conditions. Each method explored in this paper simplifies such task in different ways. DPM and DA project the PES onto a finite grid and explore decisions in making the path between the boundary conditions. In the EL approach  the surface is modified in several ways to smooth the relation between initial conditions and the end point of the trajectory. The NEB method reduces the original search over continuous path into considering only piece-wise linear paths, the number of pieces given by the number of images. It is this simplification that makes the NEB robust and accurate, since the total action now becomes a smooth function of the position of the images, a function that can straightforwardly be numerically minimized by gradient descent methods. 

A significant advantage of the NEB is that it can accommodate any initial positions of the images, which  speeds the convergence appreciably if a good prior guess of the LAP is provided. Other methods lack for such incorporation of prior knowledge. Finally, the resolution of the NEB for a rapidly varying surface can be adjusted locally by increasing the amount of images or spring constants, while for DPM and DA the entire grid resolution would have to be increased, giving an appreciable toll on the computational cost.

For both  analytic and  realistic potential energy surfaces  the NEB robustly produces a LAP.  In the cases studied, NEB outperforms the  EL and DA methods, and produces close results to those of the DPM with usually lower action integral. For many-dimensional tunneling, initiating the NEB method from the DPM path might be a winning strategy.

A huge advantage of the NEB over other methods is that it can efficiently and accurately estimate exit points. By exploring different initial conditions for the positions of the images which lead to distinct exit points,  one can use the NEB method to study the phenomenon of multimodal fission.
An example of such an application is shown in video~1 in the SM \cite{SM}. Whilst other methods can find a least-action trajectory for an arbitrary final point placed on the OTL, as done, e.g., in Refs.~\cite{Sadhukhan2013,Sadhukhan2014,Sadhukhan2016,Sadhukhan2017,Zhao2015,Zhao2016,Mercier2021}, they cannot  guarantee that this trajectory is stationary. All such trajectories can be gradually transformed into a stationary pathway by moving the final point along the OTL towards the exit point, see Fig.~\ref{Fig: Bifurcation}.

In this paper we also explored the minimum-energy (or static) path. We adjusted our NEB algorithm to generate MEPs, including the determination of local minima and saddle points.  The necessary conditions for an MEP to also be an LAP are discussed in  the SM \cite{SM}. Video~2  in the SM \cite{SM} illustrates the way the NEB method generates LAP and MEP.

An important contribution of this work is providing a beta release of the PyNEB package, a python suite of codes that implement the NEB algorithm  described in this paper. The package can be found in \cite{PyNEB} together with the respective documentation and code samples serving as a tutorial for its use. A comprehensive investigation into the intricacies of the numerical implementations and performance of the package itself will accompany the version $1.0$ release.

The NEB approach can  be readily paired with accelerated DFT calculations, such as the recent applications of Gaussian process  regression to PES emulation~\cite{Koistinen2017,Torres2019}. In these works, a Gaussian process is used to emulate the PES and DFT calculations are only run if the Gaussian process is uncertain as to the actual PES value. As NEB is not a grid-based method, it can sensibly be paired with a Gaussian process emulator that is updated as necessary while NEB runs. In this way, the LAP can be determined using far fewer DFT evaluations than is necessary in DPM.

The ability to determine the exit points is essential for determining fission fragment yields \cite{Sadhukhan2020,Sadhukhan2022}. The minimum action provides information on SF half-lives. In this context, the NEB method described in this paper is expected to speed up the global calculations of nuclear fission
for r-process simulations and studies of superheavy nuclei stability. 

\PRLsep

\begin{acknowledgments} 
 We are grateful to Edgard Bonilla and Stefan M. Wild  for useful comments. This work was supported by the U.S. Department of Energy under Award Numbers DOE-DE-NA0002847 (NNSA, the Stewardship Science Academic Alliances program), DE-SC0013365 (Office of Science), and DE-SC0018083 (Office of Science, NUCLEI SciDAC-4 collaboration) and by the National Science Foundation CSSI program under award number 2004601 (BAND collaboration). 
\end{acknowledgments}

E.F. and D.L.   contributed equally to this work.

\bibliography{references}

\end{document}


\title{Supplemental Material for ``Nudged elastic band approach to nuclear fission pathways"}

\author{Eric Flynn}
\affiliation{Department of Physics and Astronomy and FRIB/NSCL Laboratory, Michigan State University, East Lansing, Michigan 48824, USA}

\author{Daniel Lay}
\affiliation{Department of Physics and Astronomy and FRIB/NSCL Laboratory, Michigan State University, East Lansing, Michigan 48824, USA}

\author{Sylvester Agbemava}
\affiliation{FRIB/NSCL Laboratory, Michigan State University, East Lansing, Michigan 48824, USA}

\author{Pablo Giuliani}
\affiliation{FRIB/NSCL Laboratory, Michigan State University, East Lansing, Michigan 48824, USA}
\affiliation{Department of Statistics and Probability, Michigan State University, East Lansing, Michigan 48824, USA}

\author{Kyle Godbey}
\affiliation{FRIB/NSCL Laboratory, Michigan State University, East Lansing, Michigan 48824, USA}

\author{Witold Nazarewicz}
\affiliation{Department of Physics and Astronomy and FRIB/NSCL Laboratory, Michigan State University, East Lansing, Michigan 48824, USA}

\author{Jhilam Sadhukhan}
\affiliation{Physics Group, Variable Energy Cyclotron Centre, Kolkata 700064, India }
\affiliation{Homi Bhabha National Institute, Mumbai 400094, India }

\begin{abstract}
This supplemental material contains description of algorithms, videos, supplemental discussions
of  MEP and LAP equivalence conditions and the Euler-Lagrange method,
and supplemental figures S1-S3. 
\end{abstract}
\maketitle

\setcounter{figure}{0}
\setcounter{table}{0}
\renewcommand{\thefigure}{S\arabic{figure}}
\renewcommand{\thetable}{S\arabic{table}}
\renewcommand{\theequation}{S\arabic{equation}}
\renewcommand{\theHfigure}{Extended Data Fig. \thefigure}
\renewcommand{\theHtable}{Extended Data Table \thetable}
\renewcommand{\thesection}{S.\Roman{section}}

\section{Algorithms}
\label{algorithms}

\begin{algorithm}
\caption{Nudged Elastic Band}\label{alg:neb}
\begin{algorithmic}

\State Define start and end point
\State Initialize images along line
\For{Steps}
\For{Images}
\State Update effective force
\State Compute velocity of image
\State Translate image

\EndFor
\State Compute action
\State Check convergence
\EndFor

\end{algorithmic}
\end{algorithm}

\begin{algorithm}
\caption{Dynamic Programming Method}\label{alg:dpm}
\begin{algorithmic}

\State Define start and end points $\vec{q}_\textrm{in}$ and $\vec{q}_\textrm{fin}$
\For{$i=1,\ldots,n$}
\For{$j=1,\ldots,m$}
\State Select $\vec{q}_{i-1,j}$ to minimize the action from $\vec{q}_\textrm{in}$ to $\vec{q}_{ij}$, through $\vec{q}_{i-1,j}$
\State Copy the path to $\vec{q}_{i-1,j}$ and append $\vec{q}_{ij}$ to it

\EndFor
\EndFor

\State Append $\vec{q}_\textrm{fin}$ to all paths
\State Select the path with the least total action

\end{algorithmic}
\end{algorithm}

\begin{algorithm}
\caption{Dijkstra's Algorithm}\label{alg:djk}
\begin{algorithmic}

\State Define start point $\vec{q}_\textrm{in}$
\State Initialize distance matrix $d(\vec{q}_{ij})\to\infty$, $d(P_\textrm{in})\to0$
\State Mark all points as unvisited
\While{Any point is unvisited}
\State Select $\arg\min d(\vec{q}_{ij})$ with $\vec{q}_{ij}$ unvisited
\For{$\vec{q}_{i'j'}$ unvisited, $\vec{q}_{i'j'}$ a neighbor of $\vec{q}_{ij}$}
\State Compute tentative distance
\begin{equation}
    d'=d(\vec{q}_{ij})+\textrm{dist}(\vec{q}_{ij},\vec{q}_{i'j'})\nonumber
\end{equation}
\If{$d'<d(\vec{q}_{i'j'})$}
\State Set $d(\vec{q}_{i'j'})=d'$
\State Mark $\vec{q}_{i'j'}$ as coming from $\vec{q}_{ij}$
\EndIf

\EndFor
\State Mark $\vec{q}_{ij}$ as visited
\EndWhile

\end{algorithmic}
\end{algorithm}

\begin{algorithm}
\caption{Euler-Lagrange Method}\label{alg:ELE}
\begin{algorithmic}

\State Define a tolerance $\epsilon$, initial $\vec{q}_{\rm in}$ and final $\vec{q}_{\rm fin}$ points

\State Guess initial velocity $\dot{\vec{q}}_{\rm in}$ and solve Euler-Lagrange equations

\While{Final point on path is not within tolerance $|\vec{q}(t_f)-\vec{q}_{\rm fin}|>\epsilon$}
\State Vary $\dot{\vec{q}}_{\rm in}$

\EndWhile

\end{algorithmic}
\end{algorithm}

\section{Supplemental videos}

The supplemental videos (in .mp4 and  .mov formats) show the  evolution of the LAP as a function of the number of iteration steps in the NEB method.

\begin{itemize}
\item
Supplementary video 1 
(\href{https://pyneb.dev/assets/img/multi-modal.mov}{.mov}, 
\href{https://pyneb.dev/assets/img/multi-modal.mp4}{.mp4}) - This video shows the convergence behaviour of the LAP on the surface in Eq.~\eqbimodal, also shown in Fig. 2, as a function of iteration step. Circles indicate the positions of the NEB images. Multiple initial pathways are shown with one boundary image fixed to the global minimum while the remaining images are subject to gradient forces. The harmonic force is applied to the outer boundary image; it pushes the end point to the outer turning line, see Eq.~\eqharmforce. As the NEB algorithm iterates, the bands converge to two unique stationary paths with unique exit points. The computed action integral for each path is given in the legend. This example shows that the NEB is capable of identifying multiple stationary action paths.
\item
Supplementary video 2 
(\href{https://pyneb.dev/assets/img/asymm-camelback.mov}{.mov}, 
\href{https://pyneb.dev/assets/img/asymm-camelback.mp4}{.mp4}) - This video shows the convergence behaviour of the minimum energy path (MEP), shown in red, and the least action path (LAP), shown in purple on the asymmetric camel-back surface of Fig.~5. Circles indicate the positions of  the NEB images. A linear trajectory connecting two local minima with fixed endpoints is given as an initial guess. The MEP converges to a gradient curve that passes through critical points of the surface whilst the LAP bypasses two saddle points. 
\end{itemize}

\section{MEP and LAP equivalence conditions}
\label{LAP-MEP}

As discussed in Sec.~\secintroduction\, the MEP is obtained by computing the steepest descent line on the PES, and the LAP is determined by minimizing the collective action in the many-dimensional collective space.  We want to address the necessary conditions for an MEP to be an LAP.

Similar to \cite{quapp1984analysis} and Eq.~\MEPdef, we define the MEP as a gradient curve $\vec{\gamma}(\tau)$
satisfying the differential equation 
\begin{align}
    \dot{\vec{\gamma}} = \sigma(\tau) \vec{\nabla}V_\text{eff}(\vec{\gamma}(\tau)),
    \label{B1}
\end{align}
with boundary conditions  $\boldsymbol\gamma(0) = \vec{q}_{\textrm{in}}$ and $\vec{\gamma}(1) = \vec{q}_{\textrm{fin}}$. $\tau$ is an arbitrary monotonic parametrization of the curve, while $\dot{\vec{\gamma}}$ represents the local velocity of the trajectory. $\sigma(\tau)$ is a factor to account for the transformation of arc length $s$ in Eq.~\MEPdef\ to the parameter $\tau$ to be used in the EL Eq.~\EqELE. Solutions to Eq.~\eqref{B1} define a trajectory made by a collection of steepest descents and ascents between $\vec{q}_{\textrm{in}}$ and $\vec{q}_{\textrm{fin}}$ in terms of the parameter $\tau$.
\par 
On the other hand, the LAP trajectory $\boldsymbol{q}(\tau)$ is derived by finding a stationary point of the action integral 
\begin{align}
    S = \int_{0}^{1} \sqrt{2\Big(V_\text{eff}(\vec{q}) - E \Big)M_{\mu \nu}(\vec{q}) \dot{\vec{q}}^{\mu} \dot{\vec{q}}^{\nu}} d\tau,
    \label{B2}
\end{align}
where $M_{\mu \nu}$ is the inertia tensor, and the boundary conditions are $\boldsymbol{q}(0) = \vec{q}_{\textrm{in}}$ and $\boldsymbol{q}(1) = \vec{q}_{\textrm{fin}}$. Here and in the following, Einstein summation convention is assumed. By Beltrami's theorem, one can show that finding stationary solutions to the functional
\begin{align}
   \tilde{S} &= \int_{0}^{1} \Big(V_\text{eff}(\vec{q}) - E \Big)M_{\mu \nu}(\vec{q})\dot{\vec{q}}^{\mu} \dot{\vec{q}}^{\nu} d\tau \\
   &= \int_{0}^{1} g_{\mu \nu} \dot{\vec{q}}^{\mu} \dot{\vec{q}}^{\nu} d\tau
   \label{B3}
\end{align}
yields solutions that also fulfill the Euler Lagrangian equations for the functional \eqref{B2}. The converse of this statement is not true.
Eq.~\eqref{B3} is recognized as being in the same form as the Lagrangian for a free particle in a curved space with the metric tensor $g_{\mu \nu}$. By the stationary action principle,  one can derive the geodesic equation:
\begin{align}
    \ddot{q}^{\lambda} + \Gamma^{\lambda}_{\mu \nu} \dot{q}^{\mu} \dot{q}^{\nu} = 0,
\end{align}
where $\Gamma^{\lambda}_{\mu \nu}$ are the Christoffel symbols: 
\begin{align}
    \Gamma^{\lambda}_{\mu \nu} = \frac{1}{2} g^{\lambda \alpha} \Big(g_{\nu \alpha,\mu} + g_{ \mu \alpha, \nu} - g_{\mu \nu, \alpha}\Big).
\end{align}
The solutions represent a stationary point of both action functionals $\tilde{S}$ and $S$. For the case with constant inertia, the metric tensor has the form:
\begin{align}
    g_{\mu \nu} = 
    \begin{pmatrix}
    V_\text{eff}(\vec{q}) - E & 0 \\
    0 & V_\text{eff}(\vec{q}) - E 
    \end{pmatrix},
\end{align}
where we have assumed that $M_{\mu \nu}\propto \delta_{\mu\nu}$  and is  $\vec{q}$-independent. The coordinates $\vec q$ are re-scaled to be dimensionless. The geodesic equation in vector notation becomes
\begin{align}
    \label{Eq: Equations of Motion}
    (V_\text{eff} - E) \ddot{\vec{q}} = \frac{1}{2} |\dot{\vec{q}}|^{2} \vec{\nabla} V_\text{eff} - \big( \dot{\vec{q}} \cdot \vec{\nabla} V_\text{eff}\big) \dot{\vec{q}},
\end{align}
with the boundary condition $\vec{q}(0) = \vec{q}_{\textrm{in}}$ and $\vec{q}(1) = \vec{q}_{\textrm{fin}}$.

Suppose now that there exists a trajectory $\vec{q}(\tau)$ satisfying $ \vec{\nabla} V_\text{eff}\big[\vec{q}(\tau)\big] \propto  \dot{\vec{q}}$, i.e, the tangent vectors of the solution $\vec{q}(\tau)$ are parallel to the surface gradient $\vec{\nabla}V_\text{eff}$ evaluated along the curve. Since $\vec\nabla V_\text{eff}$ and $\dot{\vec{ q}}$ are parallel, Eq.~\eqref{Eq: Equations of Motion} would imply that $\ddot{\vec{q}} \propto \dot{\vec{q}}$ along the trajectory. This further implies that the tangent vectors of the solution $\vec{q}(\tau)$ are always in the same direction as its acceleration. From this we conclude that in order for  the MEP to also be a solution of Eq.~$\eqref{Eq: Equations of Motion}$, such trajectory $\vec{q}(\tau)$ can only be composed of straight lines. Critical points of the PES where $V_\text{eff}-E$ and $\vec{\nabla} V_\text{eff}$ both vanish are the only locations on the surface where the trajectory can bend non-smoothly and still obey Eq.~\eqref{Eq: Equations of Motion}.
\par 
These results significantly restrict the types of PESs for which an MEP can be an LAP. Such conditions on the surface can be obtained by the use of Eq.~\eqref{B1} as an ansatz for the equation of motion \ref{Eq: Equations of Motion}. We first note that, if the trajectory solves Eq.~\eqref{B1}, then:
\begin{align}
    &\ddot{\vec{\gamma}}(\tau) = \dot{\sigma} \vec{\nabla} V_\text{eff} +  \sigma \frac{\partial \vec{\nabla} V_\text{eff}}{\partial \vec{\gamma}} \frac{\partial \vec{\gamma}}{\partial \tau}, \\
    &\frac{\partial \vec{\nabla} V_\text{eff}}{\partial \vec{\gamma}} \frac{\partial \vec{\gamma}}{\partial \tau} =  H[\vec{\gamma}(\tau)] \dot{\vec{\gamma}},
\end{align}
where $H$ is the Hessian matrix of second derivatives of the PES evaluated along the gradient curve $\vec{\gamma}$. For such a trajectory, the acceleration along the curve is
\begin{align}
    \ddot{\vec{\gamma}}(\tau) = \dot{\sigma} \vec{\nabla} V_\text{eff} +  \sigma^{2} H \vec{\nabla} V_{\rm eff}.
    \label{curve-acceleration}
\end{align} 
Substituting this into the the geodesic equation we obtain
\begin{align}
    H \vec{\nabla} V_\text{eff} &= -\Big(\frac{|\vec{\nabla} V_\text{eff}|^{2}}{2(V_\text{eff} - E)} + \frac{\dot{\sigma}}{\sigma^{2}} \Big) \vec{\nabla} V_\text{eff}.
    \label{MEP_condition}
\end{align}
\begin{figure}[t!]
    \centering
    \includegraphics[width=0.5\textwidth]{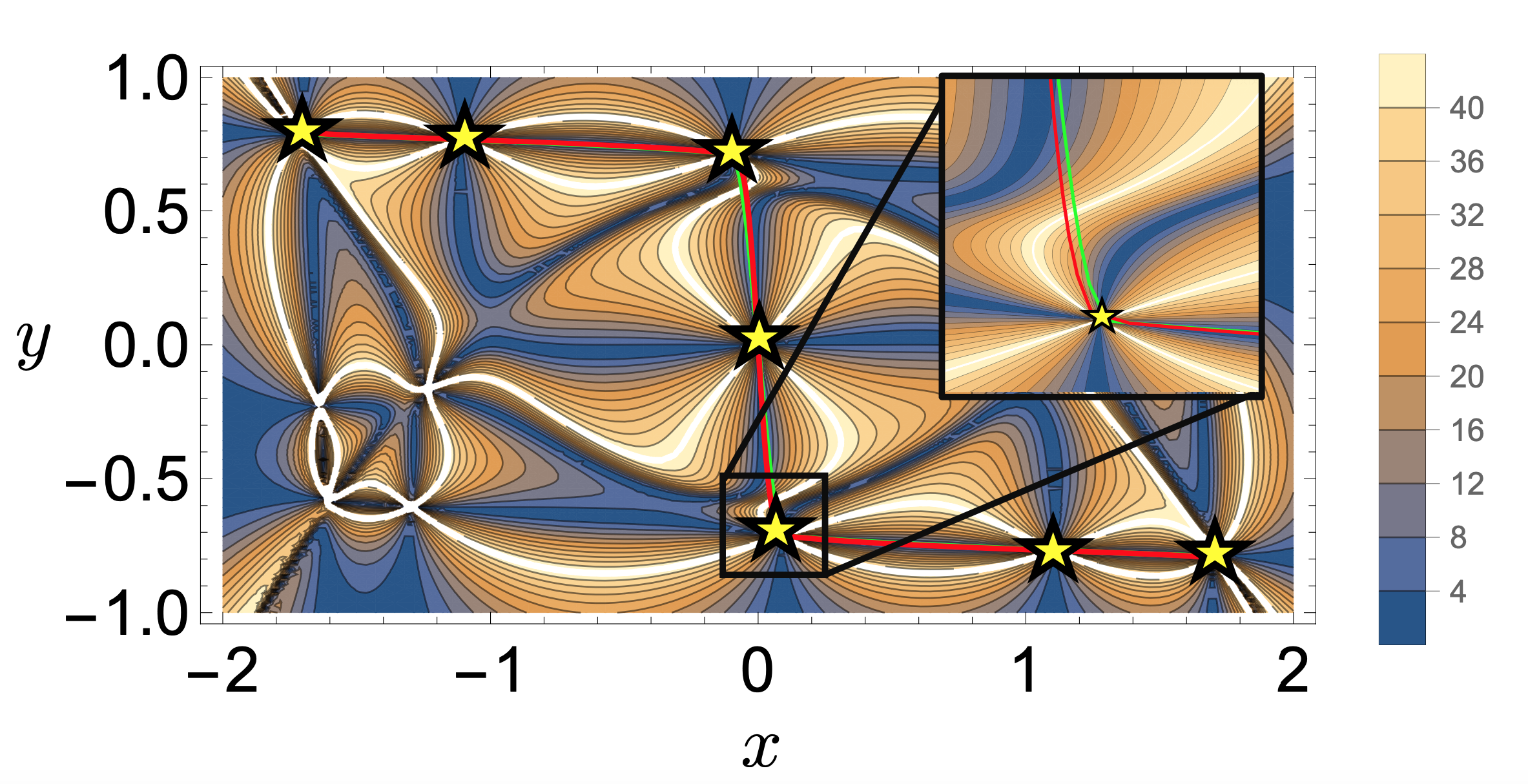}
    \caption{Absolute value of the minimum angle (in degrees) between $\vec{\nabla} V_\text{eff}$ and the eigenvectors of $H$ for the  CB-S surface of Fig.~{\figcamelsymmetric} as a function of the coordinates $(x,y)$.  The red curve shows the NEB-MEP, the green curve shows the NEB-LAP and the yellow stars indicate critical points of the surface.
    The inset shows a zoom-in on one region where there is a misalignment in the MEP path; in this region the LAP curves slightly. }
    \label{award_winning_plot}
\end{figure}
We see that the gradient $\vec{\nabla}V_\text{eff}\big[\vec{\gamma}(\tau)\big]$ must be an eigenvector of the Hessian $H\big[\vec{\gamma}(\tau)\big]$ for all $\tau$. This shows that the eigenvectors of $H$, or principal directions of the surface $V_\text{eff}$, must also be parallel to the gradient curve tangents along the trajectory. Since we know that if $\vec{\gamma}(\tau)$ is to be a solution of \eqref{Eq: Equations of Motion}, then $\dot{\vec{\gamma}}(\tau) \propto \ddot{\vec{\gamma}}(\tau)$; again showing that the curve $\vec{\gamma}(\tau)$ must be a straight line. In regions where the gradient curve is not parallel to the Hessian eigenvectors, there will be an acceleration along the gradient curve perpendicular to its tangents. In this situation, the function $\vec{\gamma}(\tau)$ cannot be a solution to \eqref{Eq: Equations of Motion} since $\ddot{\vec{q}}(\tau) \not\propto \dot{\vec{q}}(\tau)$ for all $\tau$. 

Equation~\eqref{MEP_condition} allows us to identify whether an MEP is an exact stationary path by checking if the gradient descent trajectory on the surface $V_{\rm eff}$ misaligns at any point with the eigenvectors of the surface's Hessian $H$. Figure \ref{award_winning_plot} shows the smallest angle between the gradient and the Hessian's eigenvectors for the CB-S surface. In the regions where the angle is small (stationary path can be approximated by straight lines) the MEP are LAP are close.

From this analysis, we conclude that in many situations the MEP will not coincide with the LAP. However, in cases where a stationary path can be approximated by straight lines, such as in the CB-S surface for example shown in Fig.~\figcamelsymmetric, the MEP might well approximate the LAP. Since the LAP is a stationary path, small deviations from the LAP could translate into very small errors in the action value.

\section{Details on the Euler-Lagrange method}
\label{appendix c}

Numerically solving Eq.~\EqELE\ (or Eq.~\eqref{Eq: Equations of Motion} for a constant mass case) for fixed initial conditions, $\vec q(0)=\vec q_\text{in}$ and $\dot{\vec q}(0)$, is straightforward. The main challenge lies in finding the correct initial speed and direction, $|\dot{\vec q}(0)|$ and $\hat{\dot{\vec q}}(0)$, that provides a trajectory that ends on the fixed final point $\vec q(t_f)=\vec q_\text{fin}$. In practice, it is only necessary to obtain an upper value of the initial speed $|\dot{\vec q}(0)|$ since the geometry of the trajectory would be the same with the ``particle'' arriving at an earlier time than $t_f$ to $\vec q_\text{fin}$. In the following, we discuss three important elements of the EL method implementation:
\begin{enumerate}[label=(\roman*)]
    \item How to explore the initial velocity space; 
    \item How to deal with regions where the Lagrangian in Eq.~\EqELEAction\ is close to $0$; 
    \item How to deal with boundaries (non-holonomic constraints) of the form $q_i\geq 0$. 
\end{enumerate}

With respect to (i), the main issue is that the map between $\dot{\vec q}(0)$ and $\vec q(t_f)$ is highly non-linear, and it is not trivial to create a cost function that can tell us how to improve an initial condition $\dot{\vec q}(0)$ such that $\vec q(t_f)$ gets closer to $\vec q_\text{fin}$. An automatic algorithm, such as Mathematica's shooting method \cite{Shooting}, is able to find the correct initial conditions for the analytic surfaces discussed in Sec.~\SecAnalyticSurfaces, but fails with the realistic cases of Sec.~\SecRealistic. For these realistic cases we resorted to a binary search for the initial angles of $\hat{\dot{\vec q}}(0)$. This procedure worked well for the $^{232}$U case, but yielded non-optimal results (compared to the NEB and DP paths) for the 3D $^{240}$Pu case.

The nonlinearity of $\vec q(t_f)$ as a function of $\dot{\vec q}(0)$ can get stronger 
in the regions where
the effective potential and the inertia tensor in Eq.~\actionintegrand\ undergo abrupt {local} variations. Such variations can arise due to numerical noise in the underlying DFT calculation, interpolation errors, or configuration changes due to level crossings.  The local fluctuations  can easily deflect trajectories being computed with the EL, even though their presence should not in principle impact much the actual path of minimum action.

\begin{figure}[t!]
    \includegraphics[width=0.90\linewidth]{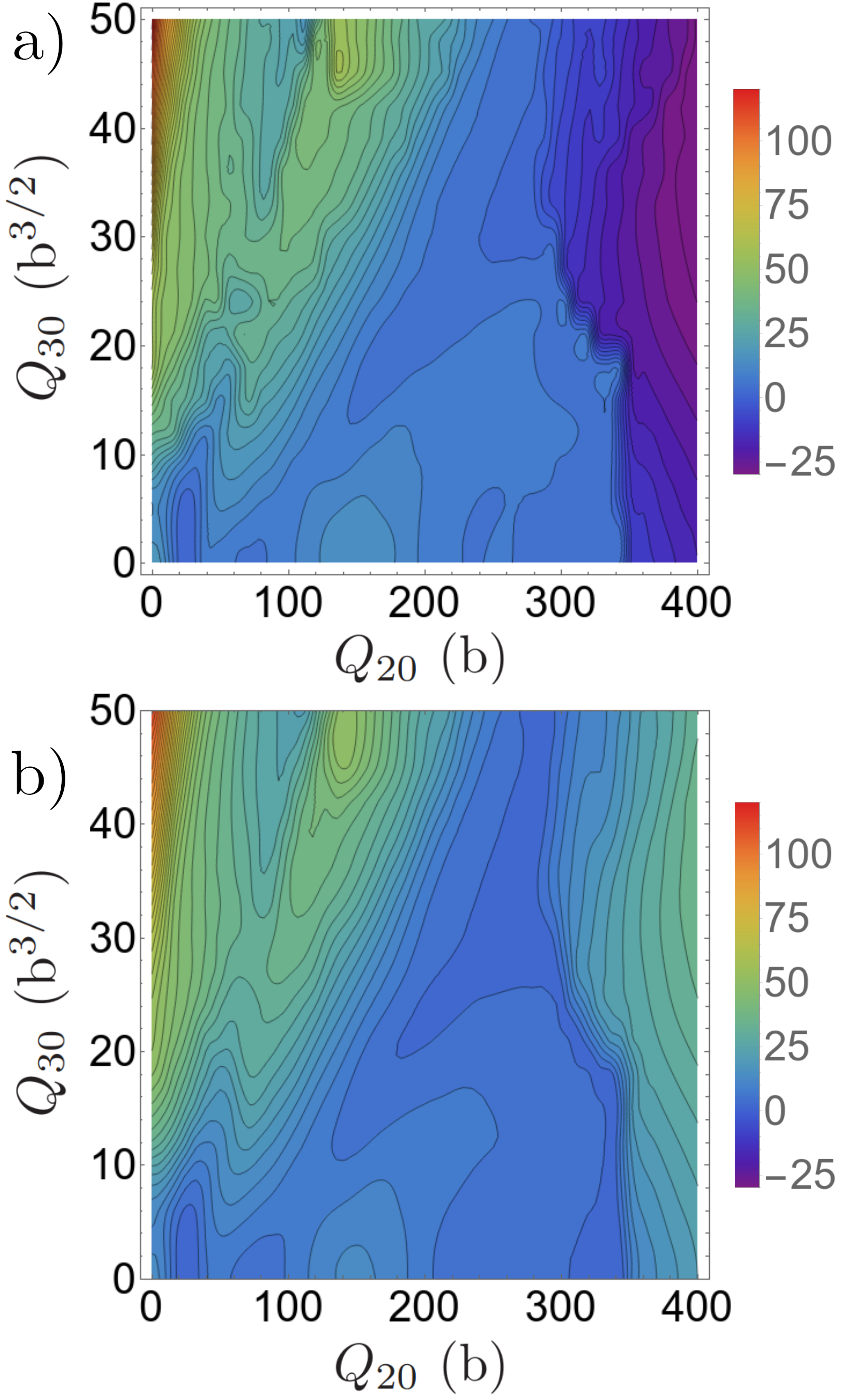}
    \caption{$^{232}$U PES before (a) and after (b) the smoothing procedure. The values near the boundary at $Q_{30}=0$ b have also been altered to guarantee that the vertical component of $\vec{\nabla} V_\text{eff}$ is zero. A similar effect is achieved in the outer turning line by replacing $(V_\text{eff}-E_0)\to \sqrt{(V_\text{eff}-E_0)^2+\epsilon_2^2}$. }
    \label{Fig: averaged surface}
\end{figure}

To remedy this situation for the EL approach, we ``smoothed'' (coarse-grained) $V_\text{eff}$ and $\mathcal{M}_{\text{eff}}$ by replacing their values at each location $\vec q$ by their respective averages over a small area around $\vec q$, see Fig.~\ref{Fig: averaged surface}. This was done at the level of the grid points, before building the interpolator. The hope is that, by removing fine-grained details of the surface, the final optimal path would not change much, but the correct $\dot{\vec q}(0)$ will be easier to find. This was not necessary for the analytic surfaces of Sec.~\SecAnalyticSurfaces\.

As concerns (ii), the Lagrangian $\mathcal{L}$ in Eq.~\EqELEAction\ becoming close to zero is a problem since the numerical integrator for the differential equations becomes unstable; increasingly big steps in $dt$ leave the action almost unchanged. This can happen if either $\big(V_\text{eff}-E_0 \big)$ or $\mathcal{M}_\text{eff}$ approach zero.

In the case of a constant inertia tensor, such as in the analytical surfaces in Sec.~\SecAnalyticSurfaces, the equations of motion Eq.~\eqref{Eq: Equations of Motion} resemble a particle of ``mass'' $\big(V_\text{eff}(\vec q)-E_0 \big)$ moving under non-conservative external forces proportional to $\nabla V_\text{eff}$ (this equation produces the same trajectories as the equations of motion of a particle moving on a inverted potential as those shown in \cite{Schmid1986}). If $V_\text{eff}\sim E_{0}$, the trajectory of the particle becomes very sensitive to any force, which is the reason why trajectories diverge near the outer turning line (see Figs. 3 and 4 in \cite{Schmid1986}), or near global minima such a those shown in Fig.~\figcamelsymmetric.

The fact that trajectories are unstable  around the OTL is because  the  assumption of the eikonal approximation that the phase of the wave function is  real and changes slowly breaks down. We refer the reader to \cite{Schmid1986} for more details on higher order corrections to the eikonal approximation.

\begin{figure}[htb!]
    \includegraphics[width=0.90\linewidth]{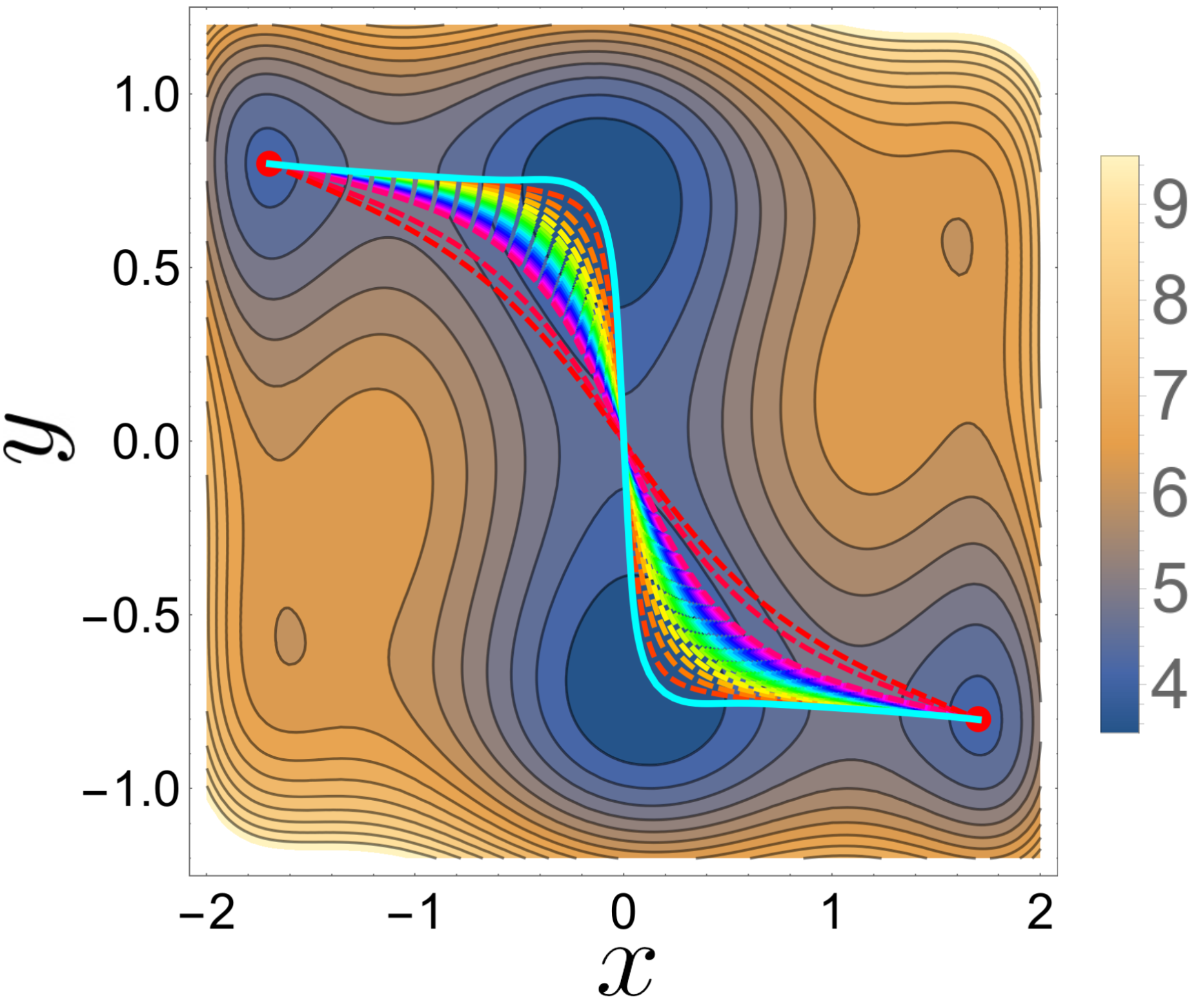}
    \caption{PES from Fig.~\figcamelsymmetric. The dashed paths are calculated with values of the potential shift in the range: $\epsilon_0 \in [0.05,3]$, while the solid cyan line is obtained by extrapolating the paths with $\epsilon_0 \in [0.05,0.42]$. The first red dashed path, $\epsilon_0=3$, has an action value around $10\%$ bigger than the last orange path, $\epsilon_0=0.05$, but it took nearly 400 times longer for the shooting method \cite{Shooting} to converge for $\epsilon_0=0.05$. This highlights how the regions where $V_\text{eff}\sim E$ lead to less stable trajectories.}
    \label{Fig: fig-camel-symmetric paths}
\end{figure}

To avoid the instabilities in regions where $V_\text{eff}\sim E_0$, including a neighborhood around the OTL, we shifted the effective potential by a small constant $\epsilon_{0}$, $V_\text{eff}-E_0 + \epsilon_0$, and added a small diagonal term to the inertia tensor: $\mathcal{M}_\text{eff} + I\epsilon_1$, where $I$ is the identity matrix with the appropriate dimensions. For the analytic surfaces in Sec.~\SecAnalyticSurfaces, we solved the EL equations for several decreasing values of $\epsilon_0$ and then extrapolated the collection of paths to obtain $\vec q(t)$ when $ \epsilon_0 \to 0$. Fig.~\ref{Fig: fig-camel-symmetric paths} shows trajectories  obtained in such way for the CB-S surface.

In the case of the realistic calculations we only worked with fixed values of $\epsilon_0$ and $\epsilon_1$. For the 2D cases we used $\epsilon_0=1$ MeV and $\epsilon_1=0.01$ ($\epsilon_1$ being dimensionless since we rescaled $Q_{20}$ and $Q_{30}$ to be in the $[0,1]$ range from their original ranges). For the 3D case we used $\epsilon_0=0.005$ MeV and $\epsilon_1=10^{-5}$.

With regard to (iii), the numerical integrator can be stopped when the trajectory arrives at  such boundary. The main issue is on how to deal with trajectories which for an appreciable interval can lay extremely close to a boundary without crossing it, such as those in Fig.~\FigUPaths\ for $Q_{20}\in \sim[25,70]$ b. For a constant effective mass, it can be seen from Eq.~\eqref{Eq: Equations of Motion} that unless $\vec\nabla V_{\rm eff}$ has zero vertical component, the particle will either be pulled down close to the boundary and touch it, or be repelled from it in the upward direction and be driven away. 

To tackle this issue, we altered the effective potential and effective mass in a small vicinity around the line $Q_{30}=0$ in such a way that the PES and effective mass values are not drastically changed, but we ensure that the vertical components of their gradients are zero along such line. This creates a ``canal effect'' with the particle  traveling close to the boundary for a finite time.

A similar ``canal effect'' was used to explore the behavior of trajectories near the OTL and improve stability of the solution in that region. This was done by replacing the effective potential by $(V_\text{eff}-E_0)\to \sqrt{(V_\text{eff}-E_0)^2+\epsilon_2^2}$, with $\epsilon_2$ being a small constant. The  gradient of this new function is zero at the OTL. In  the realistic examples, we chose the value $\epsilon_2=0.5$ MeV.

We conclude this section by emphasising that, for the EL approach to work, many simplifications and approximations had to be made. On the other hand, the NEB approach is free from the difficulties (i)-(iii).
 By using local surface derivatives and  global  springs forces, the NEB  is effectively  less sensitive to the noise in the PES. The NEB also has the advantage that the end point does not have to be predefined. For the EL approach, an undefined ending point translates into a huge computational cost.
Finally, constraints of the form $q_i\geq0$ are easily implemented on the NEB, without causing any instability of the algorithm or requiring the surface to be altered in any way.

\bibliography{references}